\documentclass[apsrev4-1,prd,twocolumn,showpacs]{revtex4-1}
\usepackage{graphicx}
\usepackage{dcolumn}
\usepackage{bm}
\usepackage{natbib}
\usepackage{multirow}
\usepackage{url}
\usepackage{color}
\usepackage{xcolor}
\usepackage{tikz}
\usepackage{float}
\usepackage[normalem]{ulem}
\usepackage{soul}

\usepackage{breakurl}

\usepackage{rotating}

\newcommand{\mbi}[1]{\mbox{\boldmath$#1$}}

\newcommand{\lsim}{\mbox{${\,\hbox{\hbox{$ < $}\kern -0.8em \lower 1.0ex\hbox{$\sim$}}\,}$}}
\newcommand{\gsim}{\mbox{${\,\hbox{\hbox{$ > $}\kern -0.8em \lower 1.0ex\hbox{$\sim$}}\,}$}}
\newcommand{\bfi}{\begin{figure}}
\newcommand{\efi}{\end{figure}}
\newcommand{\bfiw}{\begin{figure*}}
\newcommand{\efiw}{\end{figure*}}

\def\beqn{\vspace{2mm}
\begin{eqnarray}} 
\def\eeqn{\vspace{2mm} 
\end{eqnarray}}

\newcommand{\be}{\begin{equation}}
\newcommand{\ee}{\end{equation}}
\newcommand{\ba}{\begin{eqnarray}}
\newcommand{\ea}{\end{eqnarray}}
\newcommand{\brr}{\begin{array}}
 
\newcommand{\err}{\end{array}}
\newcommand{\bc}{\begin{center}}
\newcommand{\ec}{\end{center}}

\newcommand{\mnras}{MNRAS}
\newcommand{\apjs}{Rev.Astr.Astrphys.}

\newcommand{\jcap}{J.Cosm.Astr.Phys.}
\newcommand{\aap}{Astr.Astrophy.}

\newcommand{\apjl}{ApJ}

\newcommand{\aj}{AJ}

\begin{document}

%%%%%%%%%%%%%%%%%%% TITLE PAGE %%%%%%%%%%%%%%%%%%%

% Title of the paper, and the short title which is used in the headers.
% Keep the title short and informative.
\title[Linear RSD for voids]{Linear redshift space distortions for cosmic voids based on galaxies in redshift space}

% The list of authors, and the short list which is used in the headers.
% If you need two or more lines of authors, add an extra line using \newauthor
\author{Chia-Hsun Chuang$^{1*}$, Francisco-Shu Kitaura$^{1,2,3}$, Yu Liang$^{4}$, Andreu Font-Ribera$^{2,5}$, Cheng Zhao$^{4}$, Patrick McDonald$^{2}$, Charling Tao$^{4,6}$}   
%\medskip
\affiliation{
$^{1}$Leibniz-Institut f\"ur Astrophysik Potsdam (AIP), An der Sternwarte 16, D-14482 Potsdam, Germany\\
$^{2}$ Lawrence Berkeley National Lab, 1 Cyclotron Rd, Berkeley CA 94720, USA\\
$^{3}$ Departments of Physics and Astronomy, University of California, Berkeley, CA 94720, USA\\
$^{4}$ Tsinghua Center of Astrophysics and Department of Physics, Tsinghua University, Beijing 100084, China.\\
$^{5}$ Kavli IPMU (WPI), UTIAS, The University of Tokyo, Kashiwa, Chiba 277-8583, Japan
$^{6}$ Aix-Marseille Universit\'{e}, CNRS/IN2P3, CPPM UMR 7346, 13288 Marseille, France}

\email{achuang@aip.de}

\date{\today}

\begin{abstract}
Cosmic voids found in galaxy surveys are defined based on the galaxy distribution in redshift space. 
We show that the large scale distribution of voids in redshift space traces the fluctuations in the dark matter density field  $\hat{\delta}(\mbi k)$ (in Fourier space with $\mu$ being the line of sight projected $\mbi k$-vector): $\hat{\delta}_{\rm v}^s(\mbi k) = (1 + \beta_{\rm v}\mu^2)\,b^s_{\rm v}\,\hat{\delta}(\mbi k)$,  with a beta factor that will be in general different than the one describing the distribution of galaxies. Only in case voids could be assumed to be  quasi-local transformations of the linear (Gaussian)  galaxy redshift space field, one gets equal beta factors $\beta_{\rm v}=\beta_{\rm g}=f/b_{\rm g}$ with  $f$ being the growth rate, and $b_{\rm g}$, $b^s_{\rm v}$ being the galaxy and void bias on large scales defined in redshift space. 
Indeed, in our mock void catalogs we measure void beta factors being in good agreement with the galaxy one.
 Further work needs to be done to confirm the level of accuracy of the beta factor equality between voids and galaxies, but in general the void beta factor needs to be considered as a free parameter for linear RSD studies.
\end{abstract}


\maketitle

\setcounter{footnote}{0}

%%%%%%%%%%%%%%%%%%%%%%%%%%%%%%%%%%%%%%%%%%%%%%%%%%

%%%%%%%%%%%%%%%%% BODY OF PAPER %%%%%%%%%%%%%%%%%%
\section{Introduction} 
%\label{sec:intro}

Cosmic voids have drawn attention in the last few years due to their potential power to constrain cosmology and gravity. In particular, they were proposed to study the Alcock-Paczynski test (see Ref.~\citep{AP1979}), the integrated Sachs-Wolfe effect (see Ref.~\citep{Sachs1967}), weak lensing, the dark energy equation of state, modified gravity, or even the nature of dark matter   (see Refs.~\citep[][]{Granett2008,Lee2009,BPP09,Lavaux2012,Bos2012,Clampitt2013,Higuchi2013, Krause2013,Sutter2014,Cai2014a,Cai2014b,Cai2015, Lam2015,Zivick2015,Barreira2015,Clampitt2015,Yang2015,Gruenetal2016,Pollina2016,Mao2016}).
While many of these studies rely on the  shape of voids, other studies treat them as additional tracers of the density field, analogous to galaxies, or clusters of galaxies (see e.g.~Ref.~\citep[][]{W79,PP86,Bentancort1990}).
 In fact, more recently, baryon acoustic oscillations (BAO) were detected in the void clustering based on luminous red galaxies  (see Refs.~\citep{Kitauraetal2016b, Liang2016}).
The centers of voids are known to have a more linear dynamical behavior than galaxies (see Refs.~\citep[][]{SW04,Hamaus2015,Wojtak2016}).
Redshift space distortions (RSD) are interesting because they probe the growth of cosmic structures (see Ref.~\citep{Kaiser1987}) and have been successfully  studied with galaxies (see Refs.~\citep[][]{Hamilton1992, Cole1994, Cole1995, Peacocketal2001, Scoccimarro2004, Guzzo2008, Matsubara2008, Taruya2009,Percival2009, Taruya2010,Seljak2011,Jennings2011,Gil-Marin2012, Beutler2012, Reidetal2012,Blakeetal2012, Okumura2012,delaTorre2012,Valageas2013, Chuang2013, Chuang2013b, Chuang2013c, Chuang2016,delaTorreetal2013, Samushiaetal2014,Howlett2015, Reid2014,Okumura2015a, Okumura:2015lvp,Alam2015,Wang2014}).

Several recent pioneering attempts to extend RSD studies to voids have been proposed in the literature  to measure RSD from voids (see Refs.~\citep[][]{Shoji2012,Paz2013}), and to constrain the growth factor  (see Refs.~\citep[][]{Hamaus:2016wka,Cai:2016jek,Achitouv:2016mbn}).

 Voids  are not a direct observable, but are constructed based on the distribution of galaxies in redshift space.
This is a priori equivalent to a nonlinear (and nonlocal)  transformation of the density field in redshift space and introduces an additional RSD induced bias   (see Ref.~\citep[][]{Seljak2012}). 
Although one can define voids in real space from the theoretical point of view (e.g. using simulations), we actually identify voids in redshift space when analysing observations. We will show that these two definitions do not coincide.

An analogous  problem can be found in the Lyman-$\alpha$ forest (see also Ref.~\citep[][]{McDonald2000,McDonald2003,Wang2015}), for which the observable (transmitted flux fraction) is a nonlinear transformation of the quantity suffering RSD (gas density). 
We find indications, however, that in the case of voids, as long
as their arbitrary nonlinear bias involves only the linear
galaxy field in redshift space, they will share the same
beta factor, as the galaxies.
Besides Lyman-$\alpha$ forest and voids, any field constructed through a non-linear transformation applied after the effect of redshift space distortions, i.e., to a field already in redshift space (whether by physics like for the Lyman-$\alpha$ forest or through selection like voids) will have similar concerns. Generally, the standard Kaiser RSD formula relies on the field in question being conserved under the redshift space transformation, i.e., being defined in real space and simply translated into redshifted coordinates.

This paper is structured as follows, first 
we introduce the simulations used in this study and compare the measurements of correlation function with the prediction from Kaiser approximation.
Second, we consider
different bias models for cosmic voids with respect to the
galaxy field in redshift space and the relation between
the multipoles. 
In addition, we then verify our models with cross-correlation functions. Finally we present our conclusions. We show the measurements from observed data in the appendix.

\section{Measurement: multipoles of correlation functions from voids}

\bfiw
\begin{tabular}{cc}
\includegraphics[width=.42\textwidth]{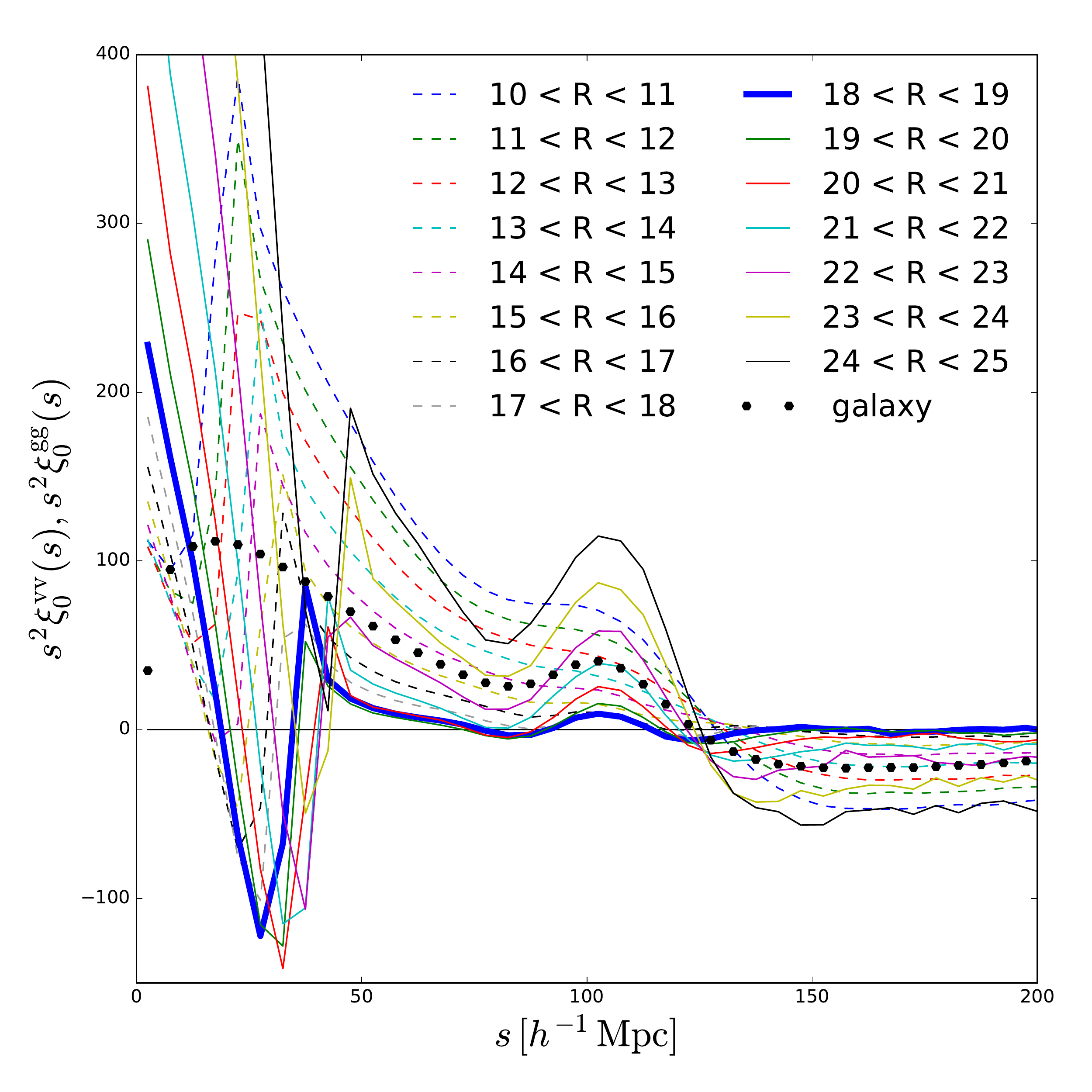}
\hspace{1cm}
\includegraphics[width=.42\textwidth]{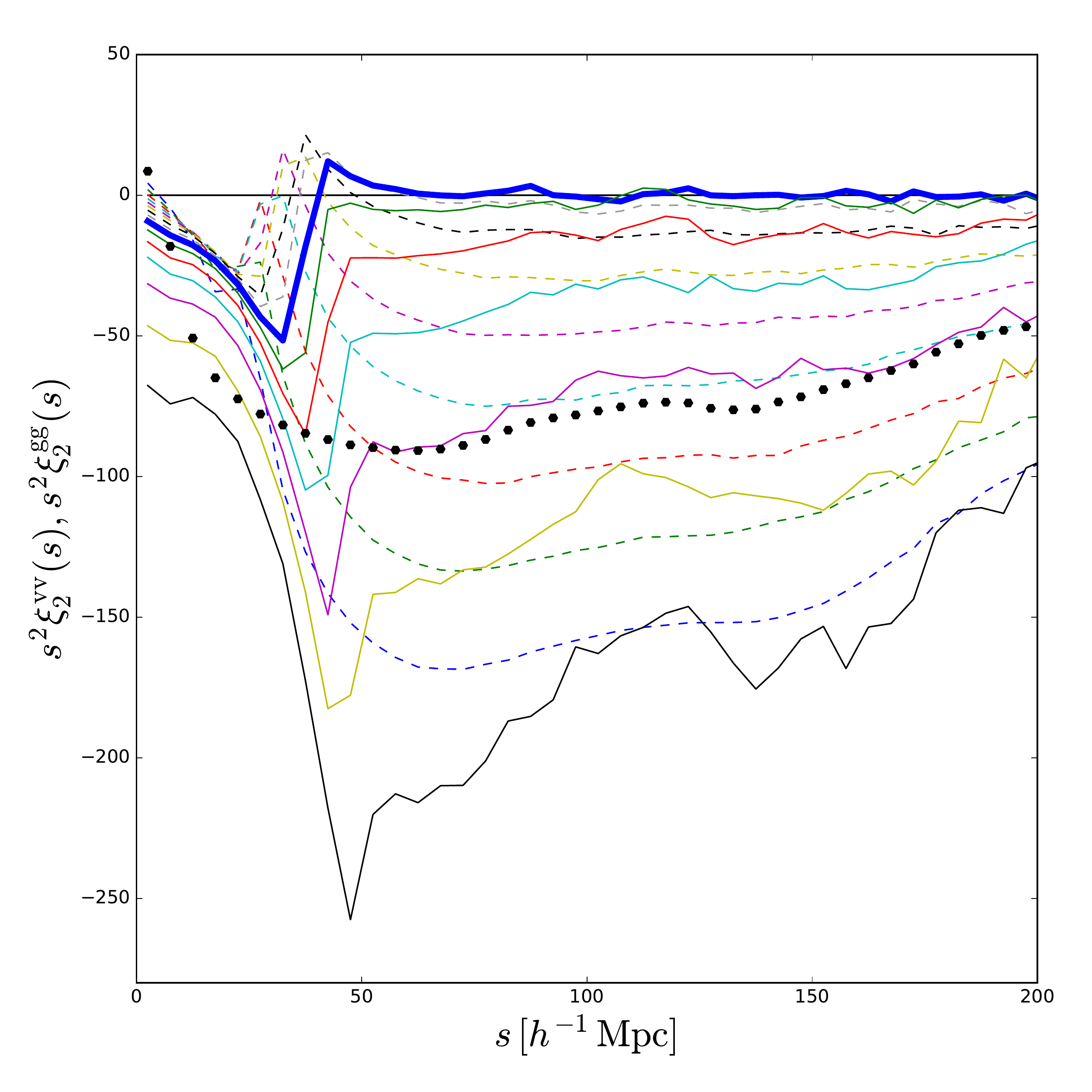}
\end{tabular}
\caption{Monopoles and quadrupoles of the auto-correlation functions measured from 100 \textsc{patchy} mock void and galaxy catalogs in boxes with different void radius $R$ bins. 
}
\label{fig:cf-mono-quad}
\efiw

\bfiw
\begin{tabular}{cc}
\includegraphics[width=.42\textwidth]{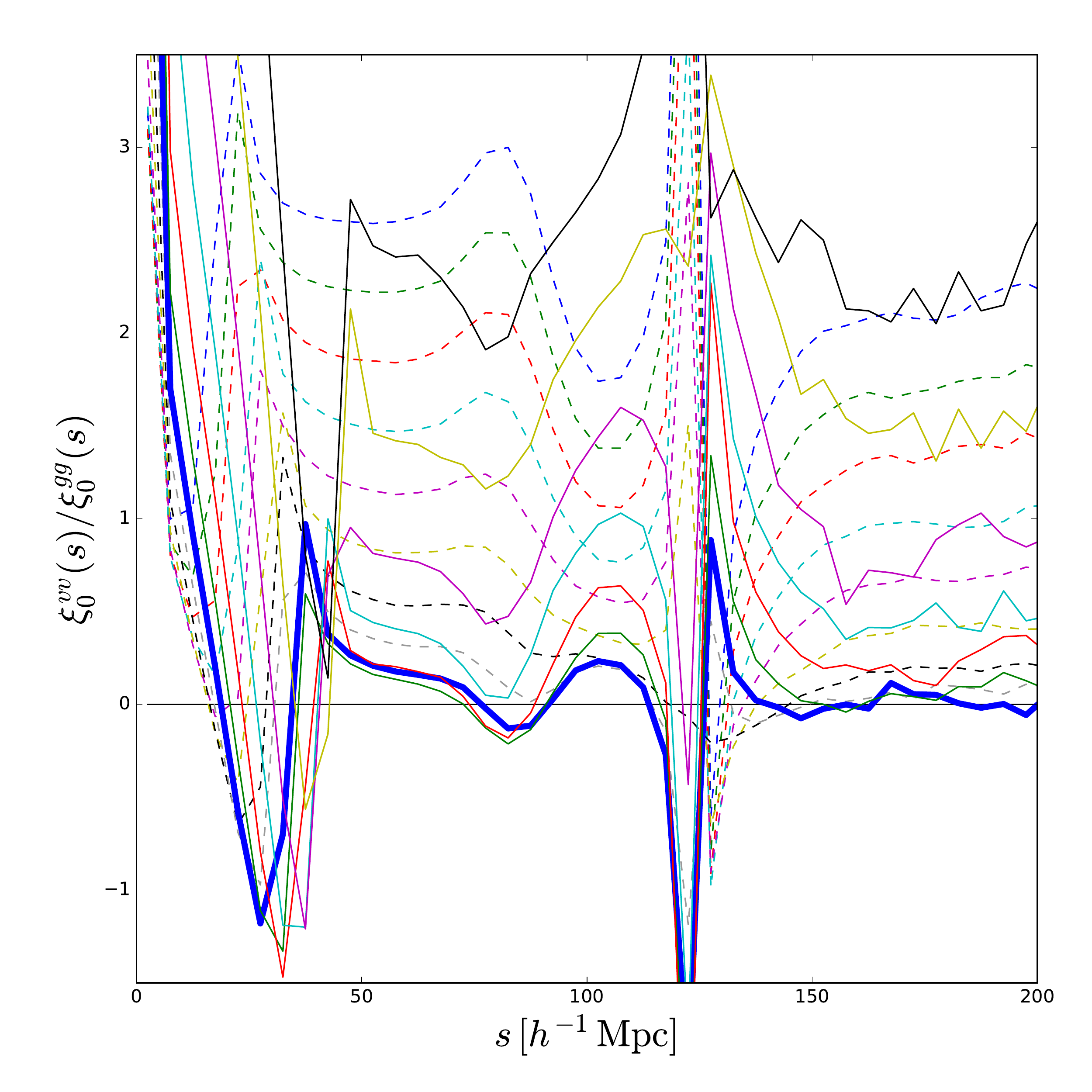}
\hspace{1cm}
\includegraphics[width=.42\textwidth]{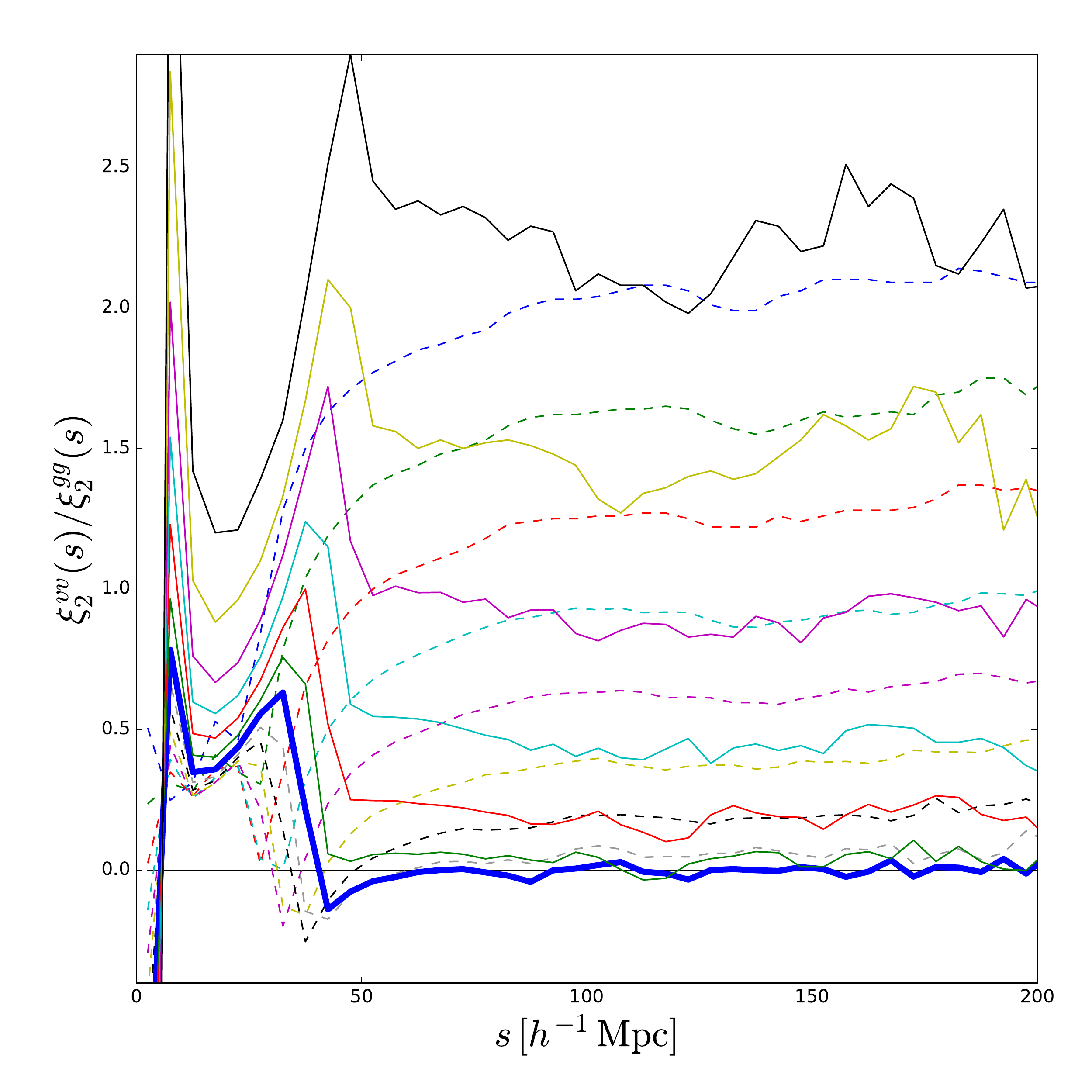}
\end{tabular}
\caption{
Using the results shown in Fig.~\ref{fig:cf-mono-quad}, we compute the ratios of the monopoles (and quadrupoles) of the voids auto-correlation functions versus the one from galaxy auto-correlation function. We compute $r_0$ and $r_2$ by averaging the scale range of $[160,200]$ $h^{-1}$ Mpc.
The color lines showing different void sizes as described in Fig.~\ref{fig:cf-mono-quad}.
}
\label{fig:void_galaxy_ratio_vs_scale}
\efiw

\bfi
\includegraphics[width=.5\textwidth]{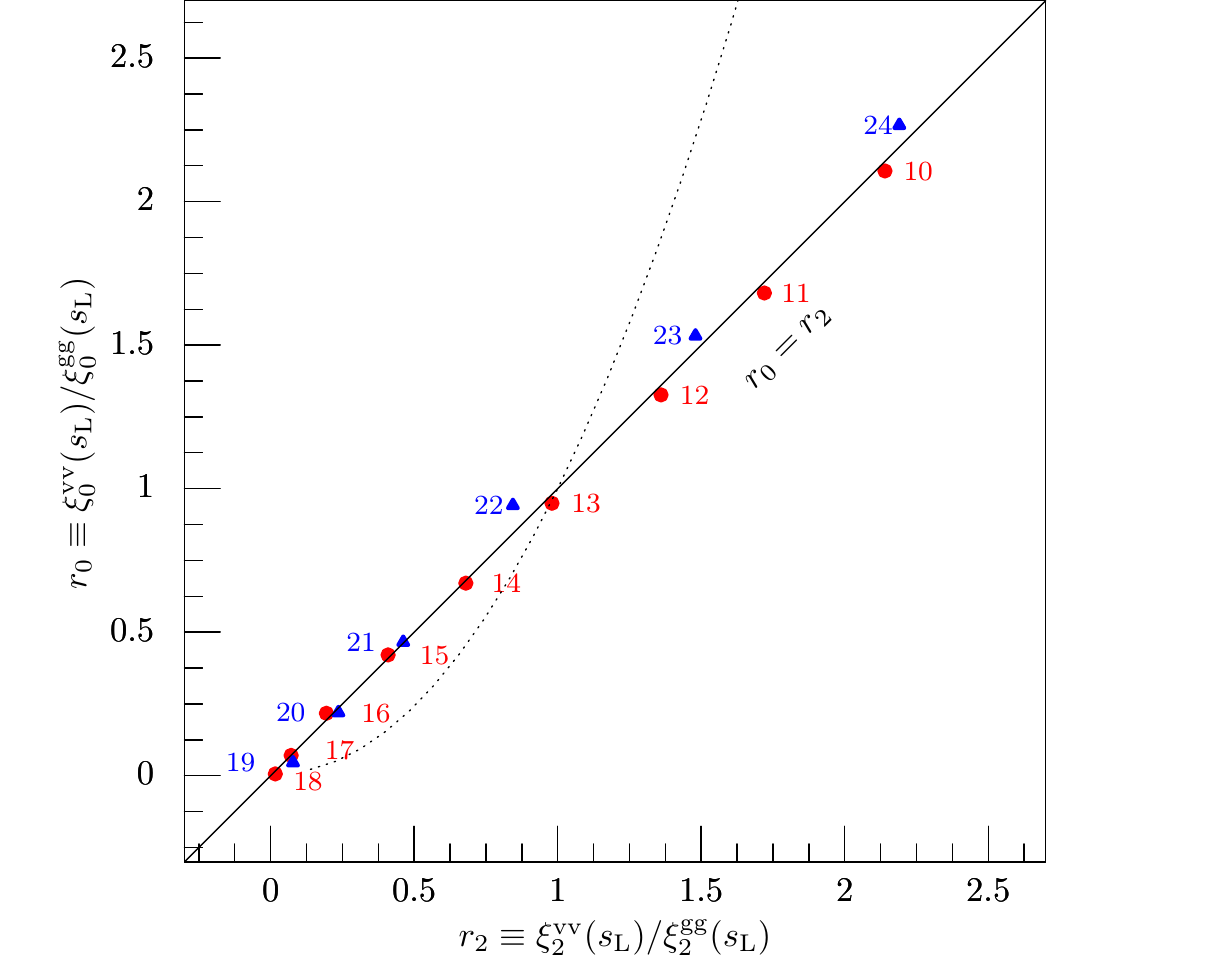}
\caption{Monopole and quadrupole ratios based on mock catalogs as shown in Fig.~\ref{fig:cf-mono-quad}. The numbers in red and blue indicate the void radius for negative and positive bias, respectively.  The void bias $b^s_{\rm vg}$ changes sign when $R$ is between 18 and 19 $h^{-1}$ Mpc.
We show also the predicion of $r_0$ vs $r_2$ from the Kaiser approximation (black dotted line). We assume the linear bias of galaxies is 2 and the growth rate at the redshift of the sample is 0.75. One can see that the prediction is very different from $r_0=r_2$ as observed from our simulations.
}
\label{fig:r_mono_r_quad}
\efi

We use 100  mock void catalogs (using the \textsc{dive} algorithm, see Ref.~\citep[][]{Zhao:2015ecx})) constructed based on mock galaxy catalogues defined in redshift space (using the \textsc{patchy} code, see Ref.~\citep[][]{Kitaura2014}), which resemble the clustering of BOSS Luminous Red Galaxies with number density around $3.5\times10^{-4}\,h^3\,\mathrm{Mpc}^{-3}$, at a mean redshift of  $z\simeq0.56$ in cubical volumes of 2.5\,$h^{-1}$Gpc side  (described in Ref.~\citep[][]{Kitauraetal2016b}).  

We compute monopoles and quadrupoles for void populations with radii ranging from 10 to 25 and bins of 1 $h^{-1}$ Mpc (see in Fig.~\ref{fig:cf-mono-quad}).

We define
\ba
r_l&\equiv&\frac{\xi^{{\rm v}{\rm v}}_l(s_{\rm L})}{\xi^{\rm g\rm g}_l(s_{\rm L})}\,,
\ea
for the different multipoles $l\in[0,2]$ of the void auto-correlation function ($\xi^{{\rm v}{\rm v}}_l$) and galaxy auto-correlation function ($\xi^{\rm g\rm g}_l$) with $s_{\rm L}$ defined on large scales. Fig.~\ref{fig:void_galaxy_ratio_vs_scale} shows the scale dependency of $r_0$ and $r_2$.

By computing the averages of the ratios within the scale range of $s_{\rm L}\in[160,200]$ $h^{-1}$ Mpc we get a very good agreement $r_0\simeq r_2$, as shown in the $r_0$-$r_2$ scatter plot for different $R$ bins on the left panel in Fig.~\ref{fig:r_mono_r_quad}. 
Note that $r_0 = r_2$ does not agree with the prediction of the Kaiser approximation as shown in  Fig.~\ref{fig:r_mono_r_quad}. Thus, a different theory, as we develop in the next section, beyond the Kaiser approximation is needed to understand what we observe in this study. We will explain in detail in the next theory section.

\section{Theory:  linear RSD for voids}

The relation between the galaxy contrast $\delta_{\rm g}$ and the dark matter field includes nonlinear, nonlocal, and stochastic components 
\citep[see, e.g.,][]{Kaiser1984,Coles1993,Fry1993,Bond1996,Dekel1999,Seljak2000,Berlind2003,McDonald2009,Desjacques2010,Matsubara2011,Schmidt2013b,Kitaura2014,Saito2014,Baldauf2015},
and can be written  for long wavelength modes as
\be
\delta_{\rm g}(\mbi r)=b_{\rm g}\delta(\mbi r)+\epsilon_{\rm g}(\mbi r)+\cdots\,,
\ee
where $b_{\rm g}$ is the linear bias, $\delta(\mbi r)$ is the dark matter field, and $\epsilon_{\rm g}$ is the galaxy noise term, followed by nonlinear and nonlocal terms.

The linear bias can be obtained from the measured clustering of galaxies, for instance the power spectrum (the auto-correlation function in Fourier space) at large scales  related to the dark matter power spectrum
\be
P_{\rm g\rm g}({\mbi k})=b_{\rm g}^2\,P({\mbi k})+P_\epsilon\,,
\ee
with $P({\mbi k})\equiv\langle\overline{\hat{\delta}(\mbi k)}\hat{\delta}(\mbi k)\rangle$, the dark matter density contrast in Fourier space given by $\hat{\delta}(\mbi k)$, and $P_\epsilon$ standing for the  noise power spectrum.

The action of gravity on large scales causes coherent flows in which galaxies tend to infall into larger density regions contributing to increment the density. This effect produces an enhancement of the power on large scales given by the Kaiser factor (see Ref.~\citep{Kaiser1987}).
Therefore,  in redshift space, the galaxy density contrast to linear order is given by 
\ba
\hat{\delta}_{\rm g}^s(\mbi k) &=& \hat{\delta}_{\rm g}(\mbi k) + f\mu^2\hat{\delta}(\mbi k)+\hat{\epsilon}\,,\\
& =& \left(1 + \beta_{\rm g}\mu^2\right)\,b_{\rm g}\,\hat{\delta}(\mbi k)+\hat{\epsilon}\,,
\ea
with $f$ being the logarithmic growth rate, $\beta_{\rm g}\equiv{f}/{b_{\rm g}}$, $\mu = \frac{\mbi k}{k} \cdot \hat{\mbi r}$, and $\hat{\mbi r}$ being the line-of-sight direction.
We will refer to the redshift space term $f\mu^2\hat{\delta}(\mbi k)$ as $\hat{\eta}_{\rm g}(\mbi k)$ in Fourier space and ${\eta}_{\rm g}(\mbi r)$ in configuration space.
Therefore the effective bias relating the galaxy density contrast in redshift space to the dark matter field can be considered to be given by $b^s_{\rm g} \equiv \left(1 + \beta_{\rm g}\mu^2\right)\,b_{\rm g} = b_{\rm g} + f\mu^2$. 
This implies that in this model the bias contribution from RSD is the same as for the dark matter (which is the unbiased case  $b_{\rm g}=1$). 
However, in general this is not true, so that a  tracer resulting from a nonlinear transformation of the density field ${\rm T}(\hat{\delta})$ with linear bias $b^{\delta}_{\rm T}$ will introduce a bias in the RSD term $b^\eta_{\rm T}$  (see Refs.~\citep[][]{Seljak2012,McDonald2000,McDonald2003,Wang2015})
\be
 b_{\rm T}(\mu) = b^{\delta}_{\rm T} + b^\eta_{\rm T}\,(f\mu^{2}) \,.
 \label{eq:bf}
\ee
where $b_{\rm T}^{\delta}$ and $b_{\rm T}^\eta$ are related to the response of the tracer ${\rm T}$ to small variations of the density and of the line-of-sight velocity gradient, respectively. 
The $b_{\rm T}^\eta$ factor is ``one'' for galaxies, as their number density is conserved in the real- to redshift-space mapping. This is however, not the case for the Lyman alpha forest or for voids.
In fact, some voids disappear or change their size in this mapping procedure  (see Ref.~\citep[][]{Zhao:2015ecx}).

We must be thus careful when constructing the bias model for voids, as these are  equivalent to a nonlinear and nonlocal transformation of the galaxy density field in redshift space.

Voids can be considered to be  tracers over an extended region characterized by their radius $R$.
Following \citet[][]{McDonald2009}, assuming isotropy and a general short-range non-locality (SRNL) kernel $K$, with the only condition that it must fall to zero outside a typical scale $R$, we can make a Taylor expansion around $\Delta \mbi r=\mbi r-\mbi r'$, to find a general expression for the void density contrast in redshift space as a function of the linear galaxy field in redshift space after considering only the leading order term
\ba
\label{eq:kernel}
\lefteqn{\delta_{\rm v}^s\left(\mbi r\right)=
\int d\Delta\mbi r K\left(\left|\Delta \mbi r\right|\right)
\delta_{\rm g}^s(\mbi r+\Delta \mbi r ) +\epsilon_{\rm v}(\mbi r)}\\
&&= \int d\Delta\mbi r K(\left|\Delta \mbi r\right|)\left[\delta_{\rm g}^s(\mbi r)+\frac{d\delta_{\rm g}^s(\mbi r)}{dr_i}\Delta r_i \right.\nonumber\\ 
&&\hspace{3.cm}\left.+\frac{1}{2}\frac{d^2\delta_{\rm g}^s(\mbi r)}{dr_i dr_j}\Delta r_i\Delta r_j
+...\right]+\epsilon_{\rm v}(\mbi r)\nonumber\\ 
&&=\delta_{\rm g}^s(\mbi r) \int d\Delta\mbi r  K\left(\left|\Delta \mbi r\right|\right)+\frac{d\delta_{\rm g}^s(\mbi r)}{dr_i} \int d\Delta\mbi r K\left(\left|\Delta \mbi r\right|\right) \Delta r_i\nonumber\\ 
&&\hspace{1.cm}+ \frac{1}{2}\frac{d^2\delta_{\rm g}^s(\mbi r)}{dr_i dr_j} \int d\Delta\mbi r K\left(\left|\Delta \mbi r\right|\right) \Delta r_i\Delta r_j +\epsilon_{\rm v}(\mbi r)+...\,,\nonumber
\ea
where $\epsilon_{\rm v}$ is the void  noise term.
Note that we have assumed the kernel is isotropic in redshift-space coordinates, which can be made true by construction at a bare (un-renormalized) level. In general SRNL can have some radial-transverse asymmetry in redshift space.

The simple integral over $K$ in the first term is a linear bias $b^s_{\rm vg}$; while the 2nd term, integrating $K~\Delta r_i$, must be zero by the symmetry of the kernel; and the third term, integrating  $K~ \Delta r_i \Delta r_j$ must be zero by symmetry if $i\neq j$, but if  $i=j$, the integral for a  generic kernel will give a result of order $R^2$  times the simple integral over the kernel in the first term, i.e.,  the integral will give a result of order $\sim b^s_{\rm vg} R^2 \delta^K_{ij}$. 
Therefore,  one gets 
\be
\delta^s_{\rm v}\left(\mbi r\right)=b^s_{\rm vg}\left[\delta_{\rm g}\left(\mbi r\right)
+ \frac{\tilde{b}_R }{2} R^2\nabla^2\delta^s_{\rm g}\left(\mbi r\right)\right]+\epsilon_{\rm v}(\mbi r)
+...
\ee
where $\tilde{b}_R$ is of order unity (e.g., if the kernel was a Gaussian with root mean square width $R$, $\tilde{b}_R$ would be exactly 1),
which in Fourier space is written as
\be
\hat{\delta}_{\rm vg}^s\left(\mbi k\right)=b^s_{\rm vg}\left[1- \frac{\tilde{b}_R}{2} R^2 k^2 \right] \hat{\delta}^s_{\rm g}\left(\mbi k\right)+\hat{\epsilon}_{\rm v}(\mbi k) +...\,.
\ee

This model permits us to assume a linear void bias within a quasi-local approximation in the large scale limit.

Let us therefore consider  the case in which voids trace only the linear part of the galaxy field in redshift space
\ba
\hat{\delta}_{\rm v}^s(\mbi k) &=& b^s_{\rm vg}\hat{\delta}_{\rm g}^s(\mbi k)+\hat{\epsilon}_{\rm v}(\mbi k)\,,\\
 &=& b^s_{\rm vg}b_{\rm g}\hat{\delta}(\mbi k) + b^s_{\rm vg}f\mu^2\hat{\delta}(\mbi k)+\hat{\epsilon}_{\rm v}(\mbi k)\,,\\
 &=& (1 + \beta_{\rm g}\mu^2)\,b^s_{\rm vg}b_{\rm g}\,\hat{\delta}(\mbi k)+\hat{\epsilon}_{\rm v}(\mbi k)\,.
\ea
This simplified model has two interesting implications. First, that the bias induced by RSD for voids on large scales is given by $b^s_{\rm vg}$ and not ``one'' as for galaxies. Second, that the beta factor $\beta_{\rm g}$ is the same as for galaxies.
The key finding of this letter is that this formula seems to describe the results of our simulations, suggesting that the approximations that go into it, i.e., neglecting non-linear effects explored later, are valid.

In this approximation, the multipoles of void power spectra can be expressed by
\ba
P^{{\rm v}{\rm v}}_l(k)&=&(b^s_{\rm vg})^2P^{\rm g\rm g}_l(k)\,,
\ea
and the multipoles of void correlation functions  by
\ba
\xi^{{\rm v}{\rm v}}_l(s)&=&(b^s_{\rm vg})^2\xi^{\rm g\rm g}_l(s)\,,
\ea
for multipoles  $l\in[0,2,4]$.
In addition, the multipoles of void cross-power spectra can be expressed by
\ba
P^{{\rm v}\rm g}_l(k)&=&b^s_{\rm vg}\,P^{\rm g\rm g}_l(k)\,,
\ea
and the multipoles of void cross-correlation functions by
\ba
\xi^{{\rm v}\rm g}_l(s)&=&b^s_{\rm vg}\,\xi^{\rm g\rm g}_l(s)\,,
\ea
where we have neglected additional noise terms.

If we consider that voids trace nonlinear galaxy density components we can demonstrate that the beta parameter for voids is not the same as for galaxies.
Below is an existence proof but not intended to be taken literally as a prediction.

Let us consider up to second order bias in  the galaxy density contrast in redshift space and neglect nonlocal bias terms
\be
\delta^s_{\rm g}(\mbi r)=b^{(1)}_{\rm g}\,\delta(\mbi r)+b^{(2)}_{\rm g}\,(\delta^2(\mbi r)-\sigma^2)+\eta(\mbi r)+\epsilon_{\rm g}(\mbi r)\,,
\ee
with $\sigma^2\equiv\langle\delta^2(\mbi r)\rangle$, including the RSD term $\eta$.

To get an expression for  the linear bias $b_{\rm g}$ one can cross correlate the galaxy field $\delta^s_{\rm g}(\mbi r)$ with the linear density field $\delta(\mbi r)$
\be
\langle\delta(\mbi r+{\rm d}\mbi r)\delta^s_{\rm g}(\mbi r)\rangle=b^{(1)}_{\rm g}\,\langle\delta(\mbi r+{\rm d}\mbi r)\delta(\mbi r)\rangle+\langle\delta(\mbi r+ {\rm d}\mbi r)\eta(\mbi r)\rangle\,.
\ee
Since we assume that $\delta$ is Gaussian, the term $\langle\delta(\mbi r+{\rm d}\mbi r)\delta^2(\mbi r)\rangle$ vanishes.
The two remaining terms can be expressed in Fourier space as $P(\mbi k)\equiv\langle\overline{\hat{\delta}(\mbi k)}\hat{\delta}(\mbi k)\rangle$ and $f\mu^2\, P(\mbi k)=\langle\overline{\hat{\delta}(\mbi k)}\hat{\eta}(\mbi k)\rangle$ yielding hence
\be
\langle\overline{\hat{\delta}(\mbi k)}\hat{\delta}^s_{\rm g}(\mbi k)\rangle=b^{(1)}_{\rm g}\,P(\mbi k)+f\mu^2\, P(\mbi k)=(1+\beta_{\rm g}\mu^2) b^{(1)}_{\rm g}P(\mbi k)\,,
\ee
with $\beta_{\rm g}\equiv f/b^{(1)}_{\rm g}$ (in our particular formulation $b_{\rm g}=b^{(1)}_{\rm g}$, for a more general case we would need to include third order terms, see Ref.~\citep[][]{McDonald2006}).

The void density contrast in redshift space can be written to third order bias as by neglecting for the sake of simplicity the convolution kernel $K$ as
\be
\delta^s_{\rm v}(\mbi r)=b^{s(1)}_{\rm vg}\,\delta^s_{\rm g}(\mbi r)+b^{s(2)}_{\rm vg}\,((\delta^s_{\rm g}(\mbi r))^2-\sigma_s^2)+b^{s(3)}_{\rm vg}\,(\delta^s_{\rm g}(\mbi r))^3+\epsilon_{\rm v}(\mbi r)\,,
\ee
with $\sigma_s^2\equiv\langle(\delta^s_{\rm g}(\mbi r))^2\rangle$, $\epsilon_{\rm v}$ being the voids shot noise.

By cross-correlating with the dark matter density contrast up to second order we get
\ba
\lefteqn{\langle{\delta}(\mbi r+{\rm d}\mbi r){\delta}^s_{\rm v}(\mbi r)\rangle}\\
&&=b^{s(1)}_{\rm vg}\langle{\delta}(\mbi r+{\rm d}\mbi r){\delta}^s_{\rm g}(\mbi r)\rangle+b^{s(2)}_{\rm vg}\langle{\delta}(\mbi r+{\rm d}\mbi r)({\delta}^s_{\rm g}(\mbi r))^2\rangle\nonumber\\
&&=b^{s(1)}_{\rm vg}b^{(1)}_{\rm g}\langle{\delta}(\mbi r+{\rm d}\mbi r){\delta}(\mbi r)\rangle+b^{s(1)}_{\rm vg}\langle{\delta}(\mbi r+{\rm d}\mbi r){\eta}(\mbi r)\rangle\nonumber\\
&&+2b^{s(2)}_{\rm vg}b^{(1)}_{\rm g}b^{(2)}_{\rm g}\langle{\delta}(\mbi r+{\rm d}\mbi r)\delta(\mbi r)(\delta(\mbi r)^2-\sigma^2)\rangle\nonumber\\
&&+2b^{s(2)}_{\rm vg}b^{(2)}_{\rm g}\langle{\delta}(\mbi r+{\rm d}\mbi r)(\delta(\mbi r)^2-\sigma^2)\eta(\mbi r)\rangle\nonumber\,,
\ea
where  we have used that $\langle\delta(\mbi r+{\rm d}\mbi r)\epsilon_{\rm v}(\mbi r)\rangle=0$ and the fact that the expected value of terms with an odd number of Gaussian variables is zero.

Using Wick's theorem we can write this in Fourier space as
\ba
\lefteqn{\langle\overline{\hat{\delta}(\mbi k)}\hat{\delta}^s_{\rm v}(\mbi k)\rangle}\\
&&=b^{s(1)}_{\rm vg}(b^{(1)}_{\rm g}+f\mu^2)P(\mbi k)+4b^{s(2)}_{\rm vg}b^{(2)}_{\rm g}(b^{(1)}_{\rm g}\sigma^2+\sigma_{\delta\eta}^2)P(\mbi k)\nonumber\,,
\ea
where $\sigma^2_{\delta\eta}\equiv\langle{\delta}(\mbi r)\eta(\mbi r)\rangle$, i.e., the zero-lag correlation of the linear density and the gradient of the velocity field. 

We can compress the above cross correlation expression to
\ba
{\langle\overline{\hat{\delta}(\mbi k)}\hat{\delta}^s_{\rm v}(\mbi k)\rangle}=(1+\beta_{\rm v}\mu^2)b^s_{\rm v}P(\mbi k)\,,
\ea
by introducing an effective void bias
\ba
b^s_{\rm v}\equiv b^{s(1)}_{\rm vg}b^{(1)}_{\rm g}+4b^{s(2)}_{\rm vg}b^{(2)}_{\rm g}(b^{(1)}_{\rm g}\sigma^2+\sigma_{\delta\eta}^2)\,,
\ea
and defining a new void beta factor
\be
\beta_{\rm v}\equiv \frac{b^{s(1)}_{\rm vg}}{b^s_{\rm v}}f=\frac{b^{s(1)}_{\rm vg}}{b^{s(1)}_{\rm vg}b^{(1)}_{\rm g}+4b^{s(2)}_{\rm vg}b^{(2)}_{\rm g}(b^{(1)}_{\rm g}\sigma^2+\sigma_{\delta\eta}^2)}f\,.
\ee
From this equation we can see that we will only have $\beta_v=\beta_g=f/b_g^{(1)}$ in the spacial case that voids are tracing the linear galaxy redshift space field, i.e., when $b^{(2)}_{\rm g}=0$.

In fact, as long as voids trace only the linear galaxy
redshift space field, the beta parameter equality between voids
and galaxies is also ensured with more complex higher order
relations.​ If we include higher order terms in the voids galaxy
relation, up to third order, and compute its cross correlation
with the linear density field we get
\ba
\label{eq:cross}
\lefteqn{\langle{\delta}(\mbi r+{\rm d}\mbi r){\delta}^s_{\rm v}(\mbi r)\rangle}\\
&&=b^{s(1)}_{\rm vg}\langle{\delta}(\mbi r+{\rm d}\mbi r){\delta}^s_{\rm g}(\mbi r)\rangle+b^{s(3)}_{\rm vg}\langle{\delta}(\mbi r+{\rm d}\mbi r)({\delta}^s_{\rm g}(\mbi r))^3\rangle\nonumber\,,
\ea
where we have used that $\langle\delta(\mbi r+{\rm d}\mbi r)\epsilon_{\rm v}(\mbi r)\rangle=0$ and $\langle\delta(\mbi r+{\rm d}\mbi r)(\delta^s_{\rm g}(\mbi r))^2\rangle=0$, since $\delta^s_{\rm g}(\mbi r)$ is also a Gaussian field.

Expanding the second term in Eq.~\ref{eq:cross} we find
\ba
\lefteqn{\langle{\delta}(\mbi r+{\rm d}\mbi r)({\delta}^s_{\rm g}(\mbi r))^3\rangle}\\
&&=b^{(1)}_{\rm g}\langle{\delta}(\mbi r+{\rm d}\mbi r)({\delta}(\mbi r))^3\rangle+\langle{\delta}(\mbi r+{\rm d}\mbi r)({\eta}(\mbi r))^3\rangle\nonumber\\
&&+3(b^{(1)}_{\rm g})^2\langle{\delta}(\mbi r+{\rm d}\mbi r)({\delta}(\mbi r))^2\eta(\mbi r)\rangle\nonumber\\
&&+3(b^{(1)}_{\rm g})^2\langle{\delta}(\mbi r+{\rm d}\mbi r){\delta}(\mbi r)(\eta(\mbi r))^2\rangle\nonumber\\
&&+3(b^{(1)}_{\rm g})^2\langle{\delta}(\mbi r+{\rm d}\mbi r){\delta}(\mbi r)(\epsilon(\mbi r))^2\rangle\nonumber\\
&&+3(b^{(1)}_{\rm g})^2\langle{\delta}(\mbi r+{\rm d}\mbi r){\eta}(\mbi r)(\epsilon(\mbi r))^2\rangle\nonumber\,,
\ea
which in Fourier space reduces to
\ba
\lefteqn{\hspace{0cm}=3(b^{(1)}_{\rm g})^2\sigma_{\delta\delta}^2P(k)+3\sigma_{\eta\eta}f\mu^2P(k)+3(b^{(1)}_{\rm g})^2\sigma_{\delta\delta}^2f\mu^2P(k)}\nonumber\\
&&+6(b^{(1)}_{\rm g})^2\sigma_{\delta\eta}^2P(k)+3b^{(1)}_{\rm g}\sigma_{\eta\eta}^2P(k)+6b^{(1)}_{\rm g}\sigma_{\delta\eta}^2f\mu^2P(k)\nonumber\\
&&+3b^{(1)}_{\rm g}\sigma_{\epsilon\epsilon}^2P(k)+3\sigma_{\epsilon\epsilon}^2f\mu^2P(k)\nonumber\\
&&=3[(b^{(1)}_{\rm g})^3\sigma_{\delta\delta}^2+2(b^{(1)}_{\rm g})^2\sigma_{\delta\eta}^2+b^{(1)}_{\rm g}\sigma_{\eta\eta}^2+b^{(1)}_{\rm g}\sigma_{\epsilon\epsilon}^2]P(k)\nonumber\\\
&&+3[\sigma_{\eta\eta}^2+(b^{(1)}_{\rm g})^2\sigma_{\delta\delta}^2+2b^{(1)}_{\rm g}\sigma_{\delta\eta}^2+\sigma_{\epsilon\epsilon}]f\mu^2P(k)\nonumber\\
&&=3[\sigma_{\eta\eta}^2+(b^{(1)}_{\rm g})^2\sigma_{\delta\delta}^2+2b^{(1)}_{\rm g}\sigma_{\delta\eta}^2+\sigma_{\epsilon\epsilon}][b^{(1)}_{\rm g}+f\mu^2]P(k)\nonumber\,,
\ea
Combining this result with the first term of Eq.~\ref{eq:cross} we get
\be
{\overline{\langle\hat{\delta}(\mbi k)}\hat{\delta}^s_{\rm v}(\mbi k)\rangle=\tilde{b}[b^{(1)}_{\rm g}+f\mu^2]P(k)}\,,
\ee
with $\tilde{b}\equiv b^{s(1)}_{\rm vg}+3b^{s(3)}_{\rm vg}[\sigma_{\eta\eta}^2+(b^{(1)}_{\rm g})^2\sigma_{\delta\delta}^2+2b^{(1)}_{\rm g}\sigma_{\delta\eta}^2+\sigma_{\epsilon\epsilon}]$.
From this we can conclude that even a non-linear transformation up to third order of the linear galaxy redshift space will retain the same beta factor: $\beta_{\rm v}=\beta_{\rm g}=f/b^{(1)}_{\rm g}$.

\section{Validation of the  RSD void model}

\bfiw
\begin{tabular}{cc}
\includegraphics[width=.42\textwidth]{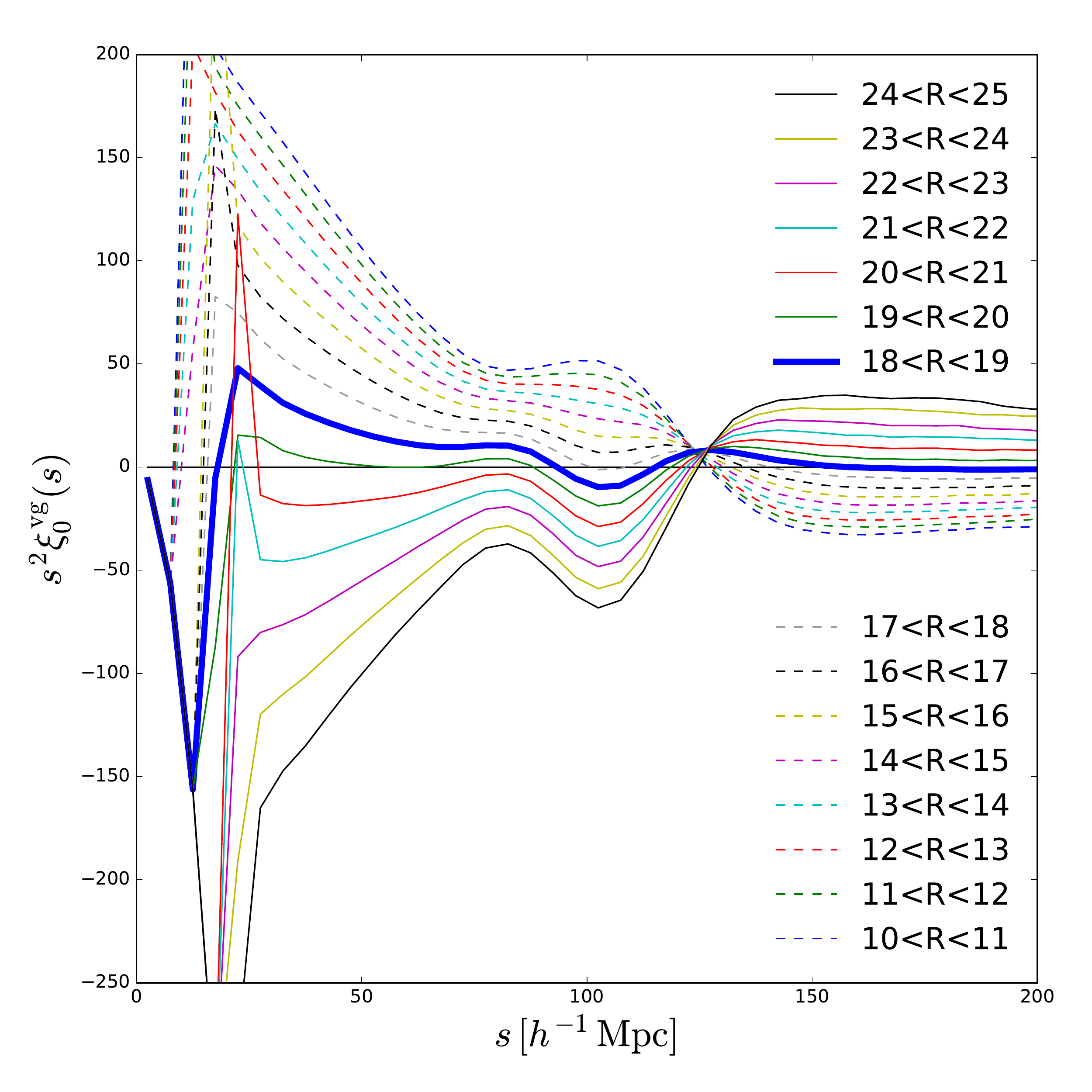}
\hspace{1cm}
\includegraphics[width=.42\textwidth]{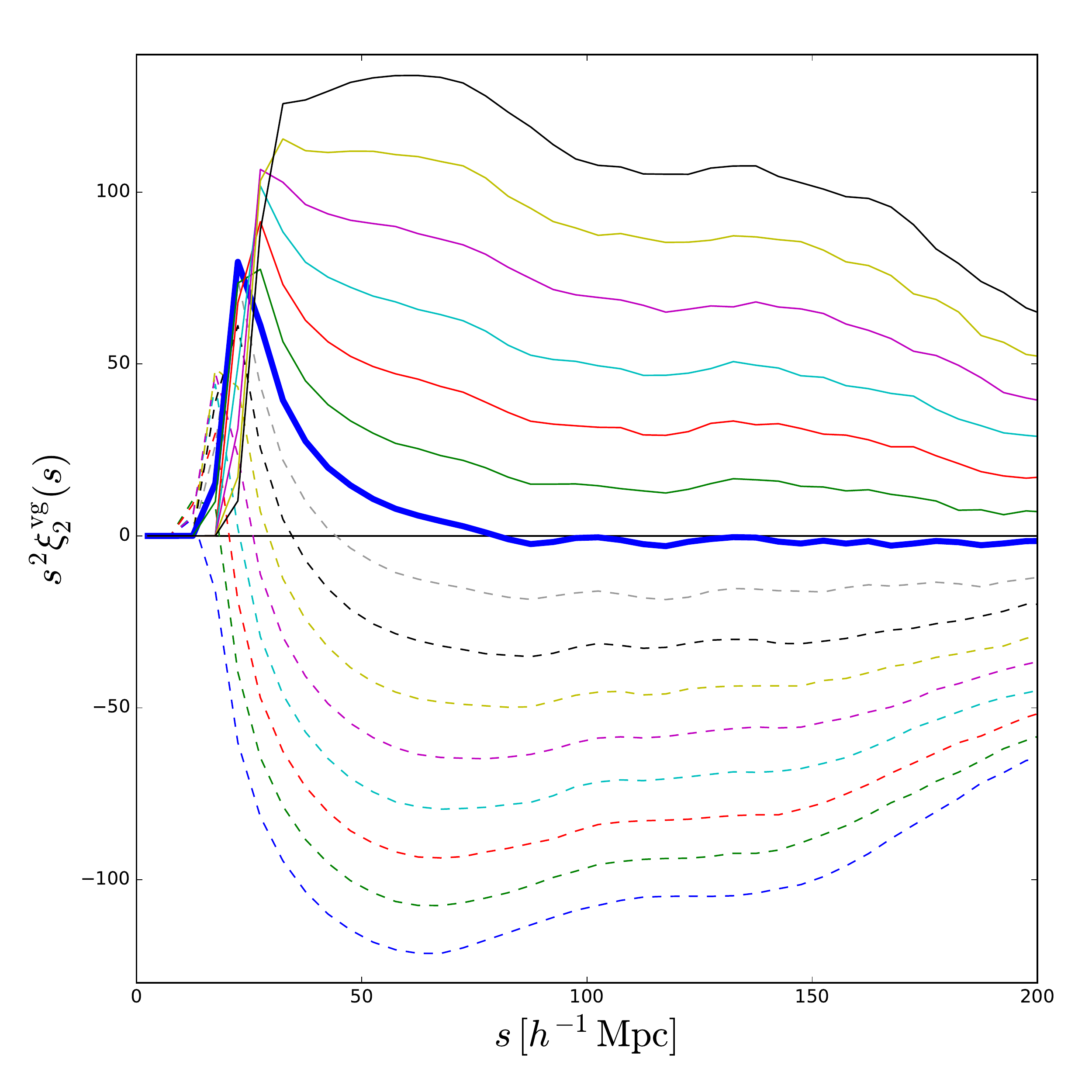}
\end{tabular}
\caption{Monopoles and quadrupoles the cross-correlation functions measured from 100 \textsc{patchy} mock void and galaxy catalogs in boxes with different void radius $R$ bins. 
}
\label{fig:xcf-mono-quad}
\efiw

\bfiw
\begin{tabular}{cc}
\includegraphics[width=.42\textwidth]{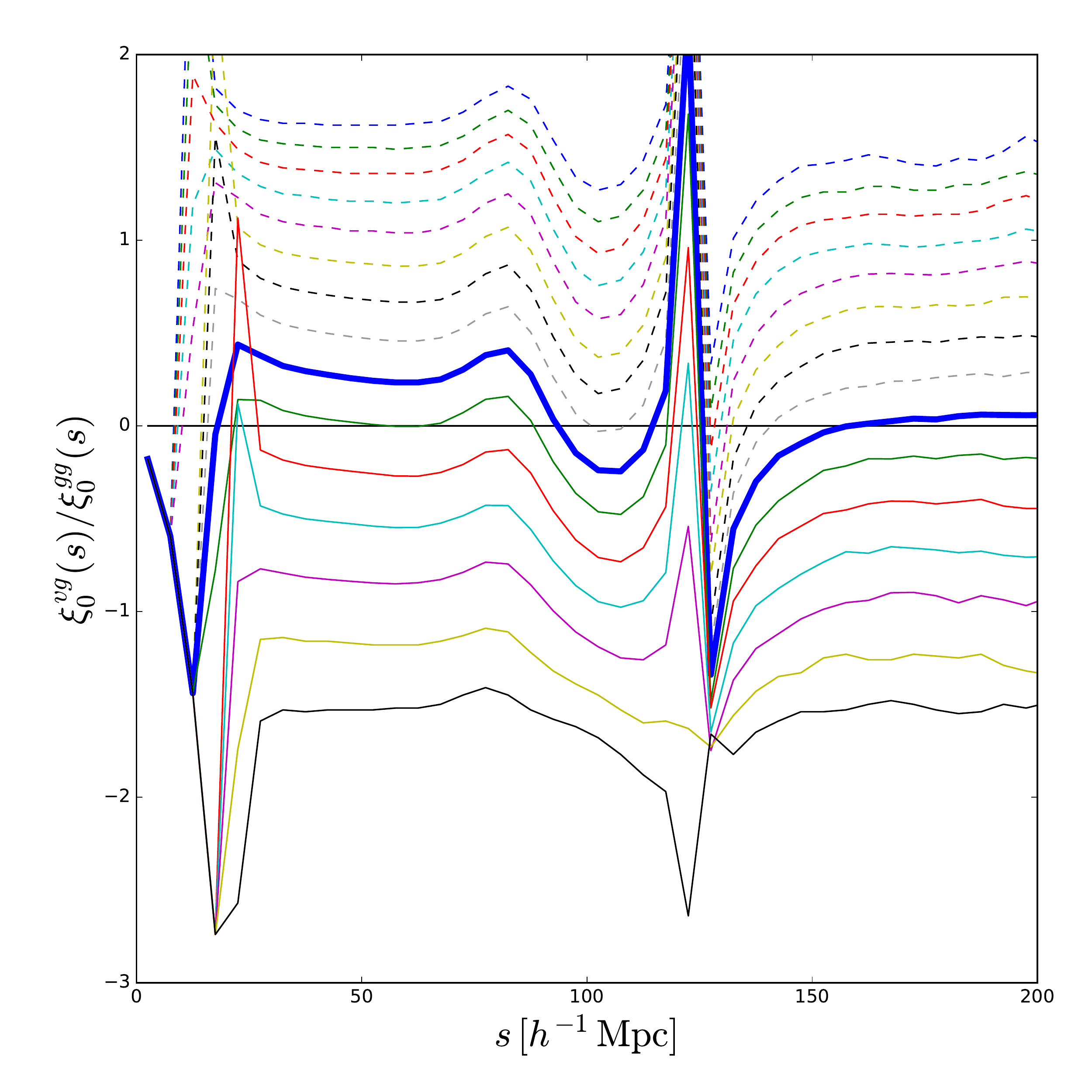}
\hspace{1cm}
\includegraphics[width=.42\textwidth]{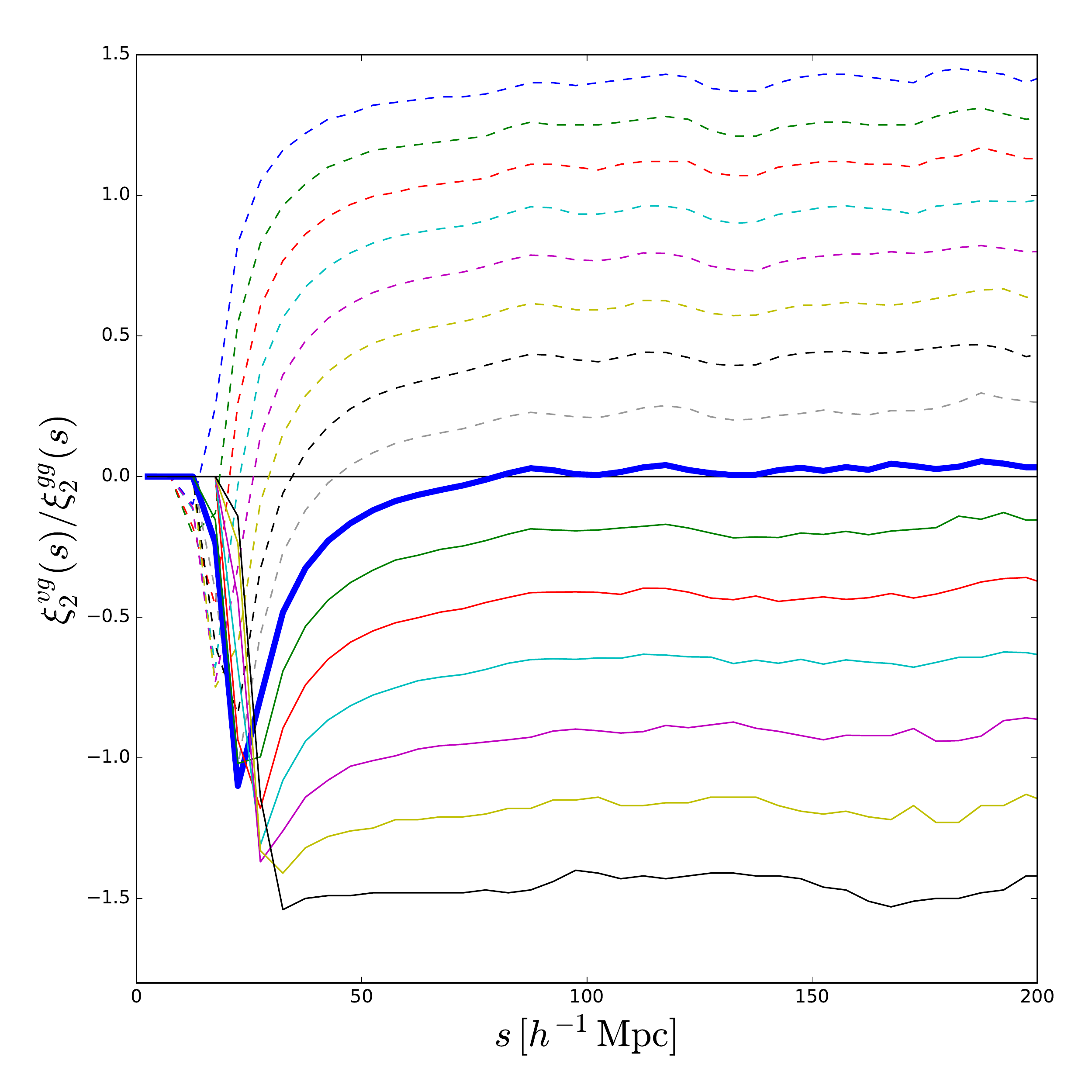}
\end{tabular}
\caption{
Using the results shown in Fig.~\ref{fig:xcf-mono-quad}, we compute the ratios of the monopoles (and quadrupoles) of the void-galaxy cross-correlation functions versus the one from galaxy auto-correlation function.
The color lines showing different void sizes as described in Fig.~\ref{fig:xcf-mono-quad}.
}
\label{fig:xcf_void_galaxy_ratio_vs_scale}
\efiw

\bfi
\includegraphics[width=.5\textwidth]{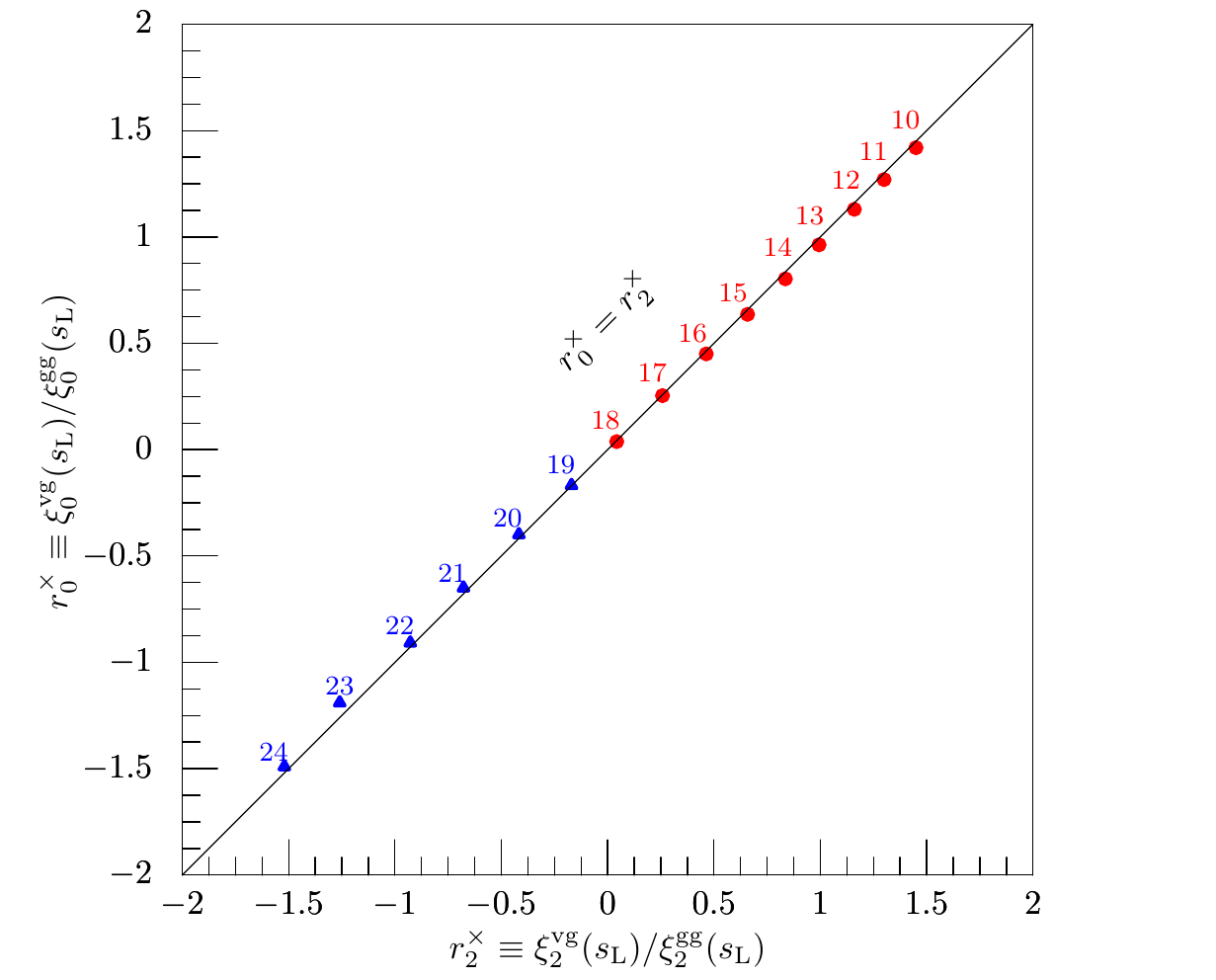}
\caption{Monopole and quadrupole ratios based on mock catalogs as shown in Fig.~\ref{fig:xcf_void_galaxy_ratio_vs_scale}. The numbers in red and blue indicate the void radius for negative and positive bias, respectively.  The void bias $b^s_{\rm vg}$ changes sign when $R$ is between 18 and 19 $h^{-1}$ Mpc.
}
\label{fig:xcf_r_mono_r_quad}
\efi

\bfi
\includegraphics[width=.5\textwidth]{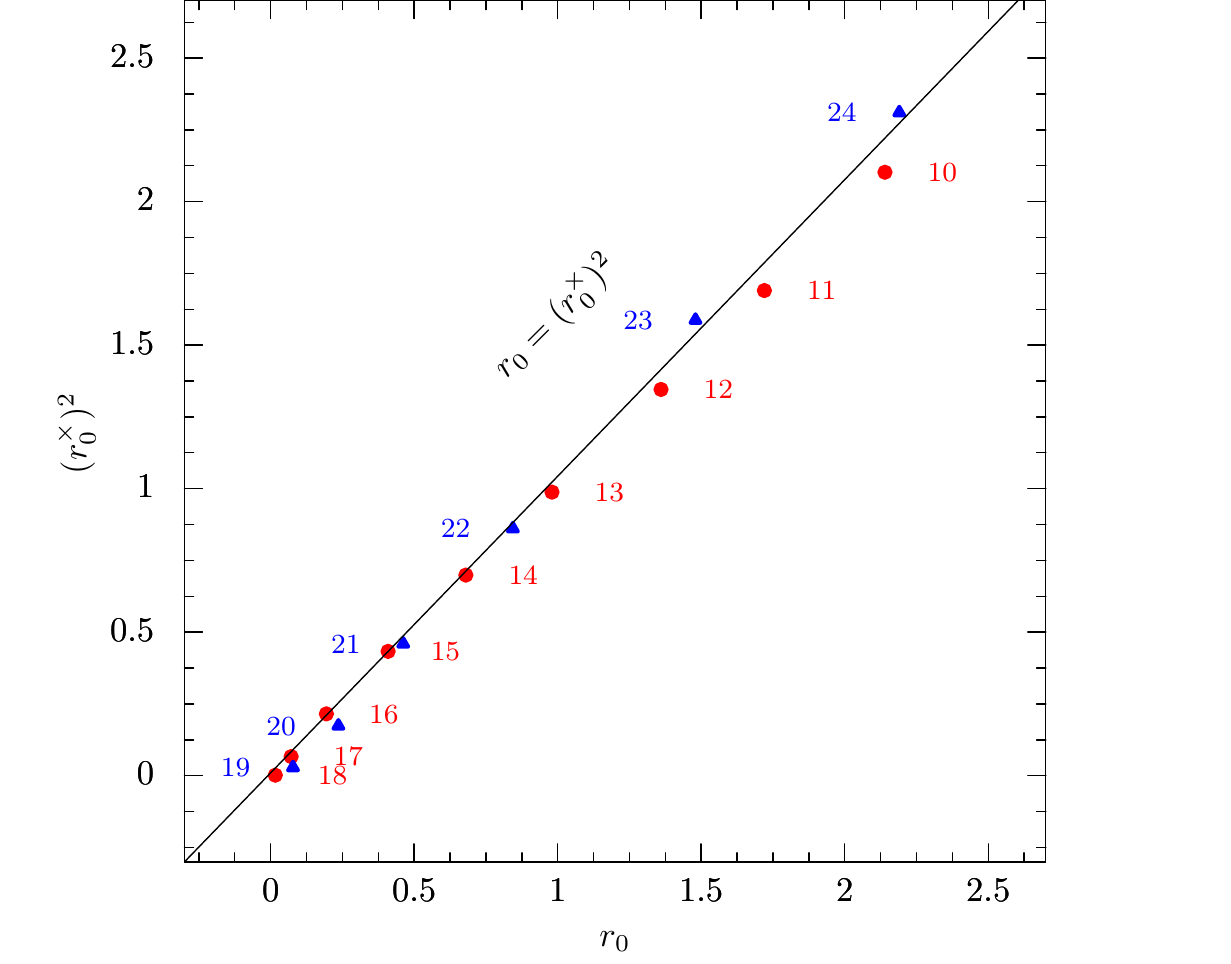}
\caption{
$(r^\times_0)^2$ vs $r_0$. The notations are the same as the ones in Fig.~\ref{fig:r_mono_r_quad} and Fig.~\ref{fig:xcf_r_mono_r_quad}. The numbers in red and blue indicate the void radius for negative and positive bias respectively. The results agree with our prediction, $(r^\times_0)^2=r_0=(b^s_{\rm vg})^2$. However, we see also slight deviation for smaller voids, e.g. $10< R < 11$ $h^{-1}$ Mpc. It should be due to the fact that smaller voids are no longer tracing only the linear galaxy density field in redshift space.
}
\label{fig:r_mono_cf_vs_xcf}
\efi

One can verify whether voids are tracing only the linear galaxy redshift space field from the multipoles of the correlation function as we have shown above, since the ratio between the void-void and the galaxy-galaxy multipoles should yield a constant value in case voids share the same beta factor as galaxies.

To reassure, we compute also the cross-correlation functions between voids and galaxeis and define the ratio between the void-galaxy and the galaxy-galaxy multipoles 
\ba
r^\times_l&\equiv&\frac{\xi^{{\rm v}{\rm g}}_l(s_{\rm L})}{\xi^{\rm g\rm g}_l(s_{\rm L})}\,,
\ea 
as shown in Fig.~\ref{fig:xcf-mono-quad} and Fig.~\ref{fig:xcf_void_galaxy_ratio_vs_scale}

For the particular  case in which voids are tracers of the linear galaxy redshift space field $r_l=(b^s_{\rm vg})^2$ and $r^\times_l=b^s_{\rm vg}$. 

The void-galaxy cross-correlation function relations lead to a very good agreement $r^{\times}_0\simeq r^{\times}_2$ as shown in 
Fig.~\ref{fig:xcf_r_mono_r_quad}.
We check also the relation between $r_0$ and $r^{\times}_0$ in Fig.~\ref{fig:r_mono_cf_vs_xcf} and find it agrees with our prediction, $(r^\times_0)^2=r_0=(b^s_{\rm vg})^2$. We see slight deviation for smaller voids, e.g. $10< R < 11$ $h^{-1}$ Mpc. It should be due to the fact that smaller voids are no longer tracing only the linear galaxy density field in redshift space.

\section{Discussion and Summary}
\label{sec:summary}

We have found that cosmic voids will in general have  a beta factor  different from the galaxy one. 
Our results based on  mock void catalogs showed void beta factors being in good agreement with the galaxy one indicating that they can be approximately assumed to be  quasi-local transformations of the linear galaxy redshift space field.

We introduced the SRNL kernel, i.e.,  Eq.~\ref{eq:kernel},
for a specific purpose: In Fig. \ref{fig:xcf-mono-quad} we see a population of
voids that appears to have zero bias in the large-separation limit; however,
they nevertheless have a BAO feature, and in fact we have found that a zero
bias population can be very good for measuring BAO \citep[][]{Kitauraetal2016b}. Since BAO
are a feature of the linear power spectrum, it is surprising that they appear
even for a zero-bias population, so this should be understood before these
voids are trusted for a distance measurement. One possibility is that the 
BAO feature comes from non-linearity, but another, probably more compelling
possibility is that we have a special case of SRNL. As we
saw above, picking a population with zero large-scale bias amounts to tuning
the integral over the SRNL kernel to be zero. However, this does not rule out,
e.g., a compensated, upside-down Mexican hat-type kernel, i.e., one that
favors the presence of a void when the density is low in the center and high
at some typical radius. The linear correlation function will then appear 
convolved with this kernel -- wherever it is smooth we will see zero, but 
where there is a feature like BAO on the scale of the kernel the correlation
will be non-zero (e.g., for a delta function feature, the result will just
look like the kernel), similar to what we see in Fig. \ref{fig:xcf-mono-quad}. 
Plots of the mean mass as a function of the
distance from void centers, which are closely related to this kernel, also look
very consistent with this understanding \cite{Zhao:2015ecx}.

\section*{Acknowledgments}

CZ, CT, and YL acknowledge support by Tsinghua University
with a 985 grant, 973 program 2013CB834906, NSFC
grant no. 11033003 and 11173017 and Sino French CNRS-CAS
international laboratories LIA Origins and FCPPL
We also thank the access to computing facilities at Barcelona (MareNostrum), at LRZ (Supermuc), at AIP (erebos),  at CCIN2P3 (Quentin Le Boulc'h), and at Tsinghua University.

%%%%%%%%%%%%%%%%%%%%%%%%%%%%%%%%%%%%%%%%%%%%%%%%%%



\bibliography{lit} % if your bibtex file is called example.bib

\begin{thebibliography}{96}%
\makeatletter
\providecommand \@ifxundefined [1]{%
 \@ifx{#1\undefined}
}%
\providecommand \@ifnum [1]{%
 \ifnum #1\expandafter \@firstoftwo
 \else \expandafter \@secondoftwo
 \fi
}%
\providecommand \@ifx [1]{%
 \ifx #1\expandafter \@firstoftwo
 \else \expandafter \@secondoftwo
 \fi
}%
\providecommand \natexlab [1]{#1}%
\providecommand \enquote  [1]{``#1''}%
\providecommand \bibnamefont  [1]{#1}%
\providecommand \bibfnamefont [1]{#1}%
\providecommand \citenamefont [1]{#1}%
\providecommand \href@noop [0]{\@secondoftwo}%
\providecommand \href [0]{\begingroup \@sanitize@url \@href}%
\providecommand \@href[1]{\@@startlink{#1}\@@href}%
\providecommand \@@href[1]{\endgroup#1\@@endlink}%
\providecommand \@sanitize@url [0]{\catcode `\\12\catcode `\$12\catcode
  `\&12\catcode `\#12\catcode `\^12\catcode `\_12\catcode `\%12\relax}%
\providecommand \@@startlink[1]{}%
\providecommand \@@endlink[0]{}%
\providecommand \url  [0]{\begingroup\@sanitize@url \@url }%
\providecommand \@url [1]{\endgroup\@href {#1}{\urlprefix }}%
\providecommand \urlprefix  [0]{URL }%
\providecommand \Eprint [0]{\href }%
\providecommand \doibase [0]{http://dx.doi.org/}%
\providecommand \selectlanguage [0]{\@gobble}%
\providecommand \bibinfo  [0]{\@secondoftwo}%
\providecommand \bibfield  [0]{\@secondoftwo}%
\providecommand \translation [1]{[#1]}%
\providecommand \BibitemOpen [0]{}%
\providecommand \bibitemStop [0]{}%
\providecommand \bibitemNoStop [0]{.\EOS\space}%
\providecommand \EOS [0]{\spacefactor3000\relax}%
\providecommand \BibitemShut  [1]{\csname bibitem#1\endcsname}%
\let\auto@bib@innerbib\@empty
%</preamble>
\bibitem [{\citenamefont {{Alcock}}\ and\ \citenamefont
  {{Paczynski}}(1979)}]{AP1979}%
  \BibitemOpen
  \bibfield  {author} {\bibinfo {author} {\bibfnamefont {C.}~\bibnamefont
  {{Alcock}}}\ and\ \bibinfo {author} {\bibfnamefont {B.}~\bibnamefont
  {{Paczynski}}},\ }\href {\doibase 10.1038/281358a0} {\bibfield  {journal}
  {\bibinfo  {journal} {\nat}\ }\textbf {\bibinfo {volume} {281}},\ \bibinfo
  {pages} {358} (\bibinfo {year} {1979})}\BibitemShut {NoStop}%
\bibitem [{\citenamefont {{Sachs}}\ and\ \citenamefont
  {{Wolfe}}(1967)}]{Sachs1967}%
  \BibitemOpen
  \bibfield  {author} {\bibinfo {author} {\bibfnamefont {R.~K.}\ \bibnamefont
  {{Sachs}}}\ and\ \bibinfo {author} {\bibfnamefont {A.~M.}\ \bibnamefont
  {{Wolfe}}},\ }\href {\doibase 10.1086/148982} {\bibfield  {journal} {\bibinfo
   {journal} {\apj}\ }\textbf {\bibinfo {volume} {147}},\ \bibinfo {pages} {73}
  (\bibinfo {year} {1967})}\BibitemShut {NoStop}%
\bibitem [{\citenamefont {{Granett}}\ \emph {et~al.}(2008)\citenamefont
  {{Granett}}, \citenamefont {{Neyrinck}},\ and\ \citenamefont
  {{Szapudi}}}]{Granett2008}%
  \BibitemOpen
  \bibfield  {author} {\bibinfo {author} {\bibfnamefont {B.~R.}\ \bibnamefont
  {{Granett}}}, \bibinfo {author} {\bibfnamefont {M.~C.}\ \bibnamefont
  {{Neyrinck}}}, \ and\ \bibinfo {author} {\bibfnamefont {I.}~\bibnamefont
  {{Szapudi}}},\ }\href {\doibase 10.1086/591670} {\bibfield  {journal}
  {\bibinfo  {journal} {\apjl}\ }\textbf {\bibinfo {volume} {683}},\ \bibinfo
  {eid} {L99} (\bibinfo {year} {2008})},\ \Eprint
  {http://arxiv.org/abs/0805.3695} {arXiv:0805.3695} \BibitemShut {NoStop}%
\bibitem [{\citenamefont {{Lee}}\ and\ \citenamefont {{Park}}(2009)}]{Lee2009}%
  \BibitemOpen
  \bibfield  {author} {\bibinfo {author} {\bibfnamefont {J.}~\bibnamefont
  {{Lee}}}\ and\ \bibinfo {author} {\bibfnamefont {D.}~\bibnamefont {{Park}}},\
  }\href {\doibase 10.1088/0004-637X/696/1/L10} {\bibfield  {journal} {\bibinfo
   {journal} {\apjl}\ }\textbf {\bibinfo {volume} {696}},\ \bibinfo {pages}
  {L10} (\bibinfo {year} {2009})},\ \Eprint {http://arxiv.org/abs/0704.0881}
  {arXiv:0704.0881} \BibitemShut {NoStop}%
\bibitem [{\citenamefont {{Betancort-Rijo}}\ \emph {et~al.}(2009)\citenamefont
  {{Betancort-Rijo}}, \citenamefont {{Patiri}}, \citenamefont {{Prada}},\ and\
  \citenamefont {{Romano}}}]{BPP09}%
  \BibitemOpen
  \bibfield  {author} {\bibinfo {author} {\bibfnamefont {J.}~\bibnamefont
  {{Betancort-Rijo}}}, \bibinfo {author} {\bibfnamefont {S.~G.}\ \bibnamefont
  {{Patiri}}}, \bibinfo {author} {\bibfnamefont {F.}~\bibnamefont {{Prada}}}, \
  and\ \bibinfo {author} {\bibfnamefont {A.~E.}\ \bibnamefont {{Romano}}},\
  }\href {\doibase 10.1111/j.1365-2966.2009.15567.x} {\bibfield  {journal}
  {\bibinfo  {journal} {\mnras}\ }\textbf {\bibinfo {volume} {400}},\ \bibinfo
  {pages} {1835} (\bibinfo {year} {2009})},\ \Eprint
  {http://arxiv.org/abs/0901.1609} {arXiv:0901.1609} \BibitemShut {NoStop}%
\bibitem [{\citenamefont {{Lavaux}}\ and\ \citenamefont
  {{Wandelt}}(2012)}]{Lavaux2012}%
  \BibitemOpen
  \bibfield  {author} {\bibinfo {author} {\bibfnamefont {G.}~\bibnamefont
  {{Lavaux}}}\ and\ \bibinfo {author} {\bibfnamefont {B.~D.}\ \bibnamefont
  {{Wandelt}}},\ }\href {\doibase 10.1088/0004-637X/754/2/109} {\bibfield
  {journal} {\bibinfo  {journal} {\apj}\ }\textbf {\bibinfo {volume} {754}},\
  \bibinfo {eid} {109} (\bibinfo {year} {2012})},\ \Eprint
  {http://arxiv.org/abs/1110.0345} {arXiv:1110.0345 [astro-ph.CO]} \BibitemShut
  {NoStop}%
\bibitem [{\citenamefont {{Bos}}\ \emph {et~al.}(2012)\citenamefont {{Bos}},
  \citenamefont {{van de Weygaert}}, \citenamefont {{Dolag}},\ and\
  \citenamefont {{Pettorino}}}]{Bos2012}%
  \BibitemOpen
  \bibfield  {author} {\bibinfo {author} {\bibfnamefont {E.~G.~P.}\
  \bibnamefont {{Bos}}}, \bibinfo {author} {\bibfnamefont {R.}~\bibnamefont
  {{van de Weygaert}}}, \bibinfo {author} {\bibfnamefont {K.}~\bibnamefont
  {{Dolag}}}, \ and\ \bibinfo {author} {\bibfnamefont {V.}~\bibnamefont
  {{Pettorino}}},\ }\href {\doibase 10.1111/j.1365-2966.2012.21478.x}
  {\bibfield  {journal} {\bibinfo  {journal} {\mnras}\ }\textbf {\bibinfo
  {volume} {426}},\ \bibinfo {pages} {440} (\bibinfo {year} {2012})},\ \Eprint
  {http://arxiv.org/abs/1205.4238} {arXiv:1205.4238 [astro-ph.CO]} \BibitemShut
  {NoStop}%
\bibitem [{\citenamefont {{Clampitt}}\ \emph {et~al.}(2013)\citenamefont
  {{Clampitt}}, \citenamefont {{Cai}},\ and\ \citenamefont
  {{Li}}}]{Clampitt2013}%
  \BibitemOpen
  \bibfield  {author} {\bibinfo {author} {\bibfnamefont {J.}~\bibnamefont
  {{Clampitt}}}, \bibinfo {author} {\bibfnamefont {Y.-C.}\ \bibnamefont
  {{Cai}}}, \ and\ \bibinfo {author} {\bibfnamefont {B.}~\bibnamefont {{Li}}},\
  }\href {\doibase 10.1093/mnras/stt219} {\bibfield  {journal} {\bibinfo
  {journal} {\mnras}\ }\textbf {\bibinfo {volume} {431}},\ \bibinfo {pages}
  {749} (\bibinfo {year} {2013})},\ \Eprint {http://arxiv.org/abs/1212.2216}
  {arXiv:1212.2216 [astro-ph.CO]} \BibitemShut {NoStop}%
\bibitem [{\citenamefont {{Higuchi}}\ \emph {et~al.}(2013)\citenamefont
  {{Higuchi}}, \citenamefont {{Oguri}},\ and\ \citenamefont
  {{Hamana}}}]{Higuchi2013}%
  \BibitemOpen
  \bibfield  {author} {\bibinfo {author} {\bibfnamefont {Y.}~\bibnamefont
  {{Higuchi}}}, \bibinfo {author} {\bibfnamefont {M.}~\bibnamefont {{Oguri}}},
  \ and\ \bibinfo {author} {\bibfnamefont {T.}~\bibnamefont {{Hamana}}},\
  }\href {\doibase 10.1093/mnras/stt521} {\bibfield  {journal} {\bibinfo
  {journal} {\mnras}\ }\textbf {\bibinfo {volume} {432}},\ \bibinfo {pages}
  {1021} (\bibinfo {year} {2013})},\ \Eprint {http://arxiv.org/abs/1211.5966}
  {arXiv:1211.5966} \BibitemShut {NoStop}%
\bibitem [{\citenamefont {{Krause}}\ \emph {et~al.}(2013)\citenamefont
  {{Krause}}, \citenamefont {{Chang}}, \citenamefont {{Dor{\'e}}},\ and\
  \citenamefont {{Umetsu}}}]{Krause2013}%
  \BibitemOpen
  \bibfield  {author} {\bibinfo {author} {\bibfnamefont {E.}~\bibnamefont
  {{Krause}}}, \bibinfo {author} {\bibfnamefont {T.-C.}\ \bibnamefont
  {{Chang}}}, \bibinfo {author} {\bibfnamefont {O.}~\bibnamefont {{Dor{\'e}}}},
  \ and\ \bibinfo {author} {\bibfnamefont {K.}~\bibnamefont {{Umetsu}}},\
  }\href {\doibase 10.1088/2041-8205/762/2/L20} {\bibfield  {journal} {\bibinfo
   {journal} {\apjl}\ }\textbf {\bibinfo {volume} {762}},\ \bibinfo {eid} {L20}
  (\bibinfo {year} {2013})},\ \Eprint {http://arxiv.org/abs/1210.2446}
  {arXiv:1210.2446} \BibitemShut {NoStop}%
\bibitem [{\citenamefont {{Sutter}}\ \emph {et~al.}(2014)\citenamefont
  {{Sutter}}, \citenamefont {{Pisani}}, \citenamefont {{Wandelt}},\ and\
  \citenamefont {{Weinberg}}}]{Sutter2014}%
  \BibitemOpen
  \bibfield  {author} {\bibinfo {author} {\bibfnamefont {P.~M.}\ \bibnamefont
  {{Sutter}}}, \bibinfo {author} {\bibfnamefont {A.}~\bibnamefont {{Pisani}}},
  \bibinfo {author} {\bibfnamefont {B.~D.}\ \bibnamefont {{Wandelt}}}, \ and\
  \bibinfo {author} {\bibfnamefont {D.~H.}\ \bibnamefont {{Weinberg}}},\ }\href
  {\doibase 10.1093/mnras/stu1392} {\bibfield  {journal} {\bibinfo  {journal}
  {\mnras}\ }\textbf {\bibinfo {volume} {443}},\ \bibinfo {pages} {2983}
  (\bibinfo {year} {2014})},\ \Eprint {http://arxiv.org/abs/1404.5618}
  {arXiv:1404.5618} \BibitemShut {NoStop}%
\bibitem [{\citenamefont {{Cai}}\ \emph
  {et~al.}(2014{\natexlab{a}})\citenamefont {{Cai}}, \citenamefont {{Li}},
  \citenamefont {{Cole}}, \citenamefont {{Frenk}},\ and\ \citenamefont
  {{Neyrinck}}}]{Cai2014a}%
  \BibitemOpen
  \bibfield  {author} {\bibinfo {author} {\bibfnamefont {Y.-C.}\ \bibnamefont
  {{Cai}}}, \bibinfo {author} {\bibfnamefont {B.}~\bibnamefont {{Li}}},
  \bibinfo {author} {\bibfnamefont {S.}~\bibnamefont {{Cole}}}, \bibinfo
  {author} {\bibfnamefont {C.~S.}\ \bibnamefont {{Frenk}}}, \ and\ \bibinfo
  {author} {\bibfnamefont {M.}~\bibnamefont {{Neyrinck}}},\ }\href {\doibase
  10.1093/mnras/stu154} {\bibfield  {journal} {\bibinfo  {journal} {\mnras}\
  }\textbf {\bibinfo {volume} {439}},\ \bibinfo {pages} {2978} (\bibinfo {year}
  {2014}{\natexlab{a}})},\ \Eprint {http://arxiv.org/abs/1310.6986}
  {arXiv:1310.6986 [astro-ph.CO]} \BibitemShut {NoStop}%
\bibitem [{\citenamefont {{Cai}}\ \emph
  {et~al.}(2014{\natexlab{b}})\citenamefont {{Cai}}, \citenamefont
  {{Neyrinck}}, \citenamefont {{Szapudi}}, \citenamefont {{Cole}},\ and\
  \citenamefont {{Frenk}}}]{Cai2014b}%
  \BibitemOpen
  \bibfield  {author} {\bibinfo {author} {\bibfnamefont {Y.-C.}\ \bibnamefont
  {{Cai}}}, \bibinfo {author} {\bibfnamefont {M.~C.}\ \bibnamefont
  {{Neyrinck}}}, \bibinfo {author} {\bibfnamefont {I.}~\bibnamefont
  {{Szapudi}}}, \bibinfo {author} {\bibfnamefont {S.}~\bibnamefont {{Cole}}}, \
  and\ \bibinfo {author} {\bibfnamefont {C.~S.}\ \bibnamefont {{Frenk}}},\
  }\href {\doibase 10.1088/0004-637X/786/2/110} {\bibfield  {journal} {\bibinfo
   {journal} {\apj}\ }\textbf {\bibinfo {volume} {786}},\ \bibinfo {eid} {110}
  (\bibinfo {year} {2014}{\natexlab{b}})},\ \Eprint
  {http://arxiv.org/abs/1301.6136} {arXiv:1301.6136} \BibitemShut {NoStop}%
\bibitem [{\citenamefont {{Cai}}\ \emph {et~al.}(2015)\citenamefont {{Cai}},
  \citenamefont {{Padilla}},\ and\ \citenamefont {{Li}}}]{Cai2015}%
  \BibitemOpen
  \bibfield  {author} {\bibinfo {author} {\bibfnamefont {Y.-C.}\ \bibnamefont
  {{Cai}}}, \bibinfo {author} {\bibfnamefont {N.}~\bibnamefont {{Padilla}}}, \
  and\ \bibinfo {author} {\bibfnamefont {B.}~\bibnamefont {{Li}}},\ }\href
  {\doibase 10.1093/mnras/stv777} {\bibfield  {journal} {\bibinfo  {journal}
  {\mnras}\ }\textbf {\bibinfo {volume} {451}},\ \bibinfo {pages} {1036}
  (\bibinfo {year} {2015})},\ \Eprint {http://arxiv.org/abs/1410.1510}
  {arXiv:1410.1510} \BibitemShut {NoStop}%
\bibitem [{\citenamefont {{Lam}}\ \emph {et~al.}(2015)\citenamefont {{Lam}},
  \citenamefont {{Clampitt}}, \citenamefont {{Cai}},\ and\ \citenamefont
  {{Li}}}]{Lam2015}%
  \BibitemOpen
  \bibfield  {author} {\bibinfo {author} {\bibfnamefont {T.~Y.}\ \bibnamefont
  {{Lam}}}, \bibinfo {author} {\bibfnamefont {J.}~\bibnamefont {{Clampitt}}},
  \bibinfo {author} {\bibfnamefont {Y.-C.}\ \bibnamefont {{Cai}}}, \ and\
  \bibinfo {author} {\bibfnamefont {B.}~\bibnamefont {{Li}}},\ }\href {\doibase
  10.1093/mnras/stv797} {\bibfield  {journal} {\bibinfo  {journal} {\mnras}\
  }\textbf {\bibinfo {volume} {450}},\ \bibinfo {pages} {3319} (\bibinfo {year}
  {2015})},\ \Eprint {http://arxiv.org/abs/1408.5338} {arXiv:1408.5338}
  \BibitemShut {NoStop}%
\bibitem [{\citenamefont {{Zivick}}\ \emph {et~al.}(2015)\citenamefont
  {{Zivick}}, \citenamefont {{Sutter}}, \citenamefont {{Wandelt}},
  \citenamefont {{Li}},\ and\ \citenamefont {{Lam}}}]{Zivick2015}%
  \BibitemOpen
  \bibfield  {author} {\bibinfo {author} {\bibfnamefont {P.}~\bibnamefont
  {{Zivick}}}, \bibinfo {author} {\bibfnamefont {P.~M.}\ \bibnamefont
  {{Sutter}}}, \bibinfo {author} {\bibfnamefont {B.~D.}\ \bibnamefont
  {{Wandelt}}}, \bibinfo {author} {\bibfnamefont {B.}~\bibnamefont {{Li}}}, \
  and\ \bibinfo {author} {\bibfnamefont {T.~Y.}\ \bibnamefont {{Lam}}},\ }\href
  {\doibase 10.1093/mnras/stv1209} {\bibfield  {journal} {\bibinfo  {journal}
  {\mnras}\ }\textbf {\bibinfo {volume} {451}},\ \bibinfo {pages} {4215}
  (\bibinfo {year} {2015})},\ \Eprint {http://arxiv.org/abs/1411.5694}
  {arXiv:1411.5694} \BibitemShut {NoStop}%
\bibitem [{\citenamefont {{Barreira}}\ \emph {et~al.}(2015)\citenamefont
  {{Barreira}}, \citenamefont {{Li}}, \citenamefont {{Jennings}}, \citenamefont
  {{Merten}}, \citenamefont {{King}}, \citenamefont {{Baugh}},\ and\
  \citenamefont {{Pascoli}}}]{Barreira2015}%
  \BibitemOpen
  \bibfield  {author} {\bibinfo {author} {\bibfnamefont {A.}~\bibnamefont
  {{Barreira}}}, \bibinfo {author} {\bibfnamefont {B.}~\bibnamefont {{Li}}},
  \bibinfo {author} {\bibfnamefont {E.}~\bibnamefont {{Jennings}}}, \bibinfo
  {author} {\bibfnamefont {J.}~\bibnamefont {{Merten}}}, \bibinfo {author}
  {\bibfnamefont {L.}~\bibnamefont {{King}}}, \bibinfo {author} {\bibfnamefont
  {C.~M.}\ \bibnamefont {{Baugh}}}, \ and\ \bibinfo {author} {\bibfnamefont
  {S.}~\bibnamefont {{Pascoli}}},\ }\href {\doibase 10.1093/mnras/stv2211}
  {\bibfield  {journal} {\bibinfo  {journal} {\mnras}\ }\textbf {\bibinfo
  {volume} {454}},\ \bibinfo {pages} {4085} (\bibinfo {year} {2015})},\ \Eprint
  {http://arxiv.org/abs/1505.03468} {arXiv:1505.03468} \BibitemShut {NoStop}%
\bibitem [{\citenamefont {{Clampitt}}\ and\ \citenamefont
  {{Jain}}(2015)}]{Clampitt2015}%
  \BibitemOpen
  \bibfield  {author} {\bibinfo {author} {\bibfnamefont {J.}~\bibnamefont
  {{Clampitt}}}\ and\ \bibinfo {author} {\bibfnamefont {B.}~\bibnamefont
  {{Jain}}},\ }\href {\doibase 10.1093/mnras/stv2215} {\bibfield  {journal}
  {\bibinfo  {journal} {\mnras}\ }\textbf {\bibinfo {volume} {454}},\ \bibinfo
  {pages} {3357} (\bibinfo {year} {2015})},\ \Eprint
  {http://arxiv.org/abs/1404.1834} {arXiv:1404.1834} \BibitemShut {NoStop}%
\bibitem [{\citenamefont {{Yang}}\ \emph {et~al.}(2015)\citenamefont {{Yang}},
  \citenamefont {{Neyrinck}}, \citenamefont {{Arag{\'o}n-Calvo}}, \citenamefont
  {{Falck}},\ and\ \citenamefont {{Silk}}}]{Yang2015}%
  \BibitemOpen
  \bibfield  {author} {\bibinfo {author} {\bibfnamefont {L.~F.}\ \bibnamefont
  {{Yang}}}, \bibinfo {author} {\bibfnamefont {M.~C.}\ \bibnamefont
  {{Neyrinck}}}, \bibinfo {author} {\bibfnamefont {M.~A.}\ \bibnamefont
  {{Arag{\'o}n-Calvo}}}, \bibinfo {author} {\bibfnamefont {B.}~\bibnamefont
  {{Falck}}}, \ and\ \bibinfo {author} {\bibfnamefont {J.}~\bibnamefont
  {{Silk}}},\ }\href {\doibase 10.1093/mnras/stv1087} {\bibfield  {journal}
  {\bibinfo  {journal} {\mnras}\ }\textbf {\bibinfo {volume} {451}},\ \bibinfo
  {pages} {3606} (\bibinfo {year} {2015})},\ \Eprint
  {http://arxiv.org/abs/1411.5029} {arXiv:1411.5029} \BibitemShut {NoStop}%
\bibitem [{\citenamefont {{Gruen et al.}}(2016)}]{Gruenetal2016}%
  \BibitemOpen
  \bibfield  {author} {\bibinfo {author} {\bibfnamefont {D.}~\bibnamefont
  {{Gruen et al.}}},\ }\href {\doibase 10.1093/mnras/stv2506} {\bibfield
  {journal} {\bibinfo  {journal} {\mnras}\ }\textbf {\bibinfo {volume} {455}},\
  \bibinfo {pages} {3367} (\bibinfo {year} {2016})},\ \Eprint
  {http://arxiv.org/abs/1507.05090} {arXiv:1507.05090} \BibitemShut {NoStop}%
\bibitem [{\citenamefont {{Pollina}}\ \emph {et~al.}(2016)\citenamefont
  {{Pollina}}, \citenamefont {{Baldi}}, \citenamefont {{Marulli}},\ and\
  \citenamefont {{Moscardini}}}]{Pollina2016}%
  \BibitemOpen
  \bibfield  {author} {\bibinfo {author} {\bibfnamefont {G.}~\bibnamefont
  {{Pollina}}}, \bibinfo {author} {\bibfnamefont {M.}~\bibnamefont {{Baldi}}},
  \bibinfo {author} {\bibfnamefont {F.}~\bibnamefont {{Marulli}}}, \ and\
  \bibinfo {author} {\bibfnamefont {L.}~\bibnamefont {{Moscardini}}},\ }\href
  {\doibase 10.1093/mnras/stv2503} {\bibfield  {journal} {\bibinfo  {journal}
  {\mnras}\ }\textbf {\bibinfo {volume} {455}},\ \bibinfo {pages} {3075}
  (\bibinfo {year} {2016})},\ \Eprint {http://arxiv.org/abs/1506.08831}
  {arXiv:1506.08831} \BibitemShut {NoStop}%
\bibitem [{\citenamefont {{Mao}}\ \emph {et~al.}(2016)\citenamefont {{Mao}},
  \citenamefont {{Berlind}}, \citenamefont {{Scherrer}}, \citenamefont
  {{Neyrinck}}, \citenamefont {{Scoccimarro}}, \citenamefont {{Tinker}},\ and\
  \citenamefont {{McBride}}}]{Mao2016}%
  \BibitemOpen
  \bibfield  {author} {\bibinfo {author} {\bibfnamefont {Q.}~\bibnamefont
  {{Mao}}}, \bibinfo {author} {\bibfnamefont {A.~A.}\ \bibnamefont
  {{Berlind}}}, \bibinfo {author} {\bibfnamefont {R.~J.}\ \bibnamefont
  {{Scherrer}}}, \bibinfo {author} {\bibfnamefont {M.~C.}\ \bibnamefont
  {{Neyrinck}}}, \bibinfo {author} {\bibfnamefont {R.}~\bibnamefont
  {{Scoccimarro}}}, \bibinfo {author} {\bibfnamefont {J.~L.}\ \bibnamefont
  {{Tinker}}}, \ and\ \bibinfo {author} {\bibfnamefont {C.~K.}\ \bibnamefont
  {{McBride}}},\ }\href@noop {} {\bibfield  {journal} {\bibinfo  {journal}
  {ArXiv e-prints}\ } (\bibinfo {year} {2016})},\ \Eprint
  {http://arxiv.org/abs/1602.06306} {arXiv:1602.06306} \BibitemShut {NoStop}%
\bibitem [{\citenamefont {{White}}(1979)}]{W79}%
  \BibitemOpen
  \bibfield  {author} {\bibinfo {author} {\bibfnamefont {S.~D.~M.}\
  \bibnamefont {{White}}},\ }\href@noop {} {\bibfield  {journal} {\bibinfo
  {journal} {\mnras}\ }\textbf {\bibinfo {volume} {186}},\ \bibinfo {pages}
  {145} (\bibinfo {year} {1979})}\BibitemShut {NoStop}%
\bibitem [{\citenamefont {{Politzer}}\ and\ \citenamefont
  {{Preskill}}(1986)}]{PP86}%
  \BibitemOpen
  \bibfield  {author} {\bibinfo {author} {\bibfnamefont {H.~D.}\ \bibnamefont
  {{Politzer}}}\ and\ \bibinfo {author} {\bibfnamefont {J.~P.}\ \bibnamefont
  {{Preskill}}},\ }\href {\doibase 10.1103/PhysRevLett.56.99} {\bibfield
  {journal} {\bibinfo  {journal} {Physical Review Letters}\ }\textbf {\bibinfo
  {volume} {56}},\ \bibinfo {pages} {99} (\bibinfo {year} {1986})}\BibitemShut
  {NoStop}%
\bibitem [{\citenamefont {{Betancort-Rijo}}(1990)}]{Bentancort1990}%
  \BibitemOpen
  \bibfield  {author} {\bibinfo {author} {\bibfnamefont {J.}~\bibnamefont
  {{Betancort-Rijo}}},\ }\href@noop {} {\bibfield  {journal} {\bibinfo
  {journal} {\mnras}\ }\textbf {\bibinfo {volume} {246}},\ \bibinfo {pages}
  {608} (\bibinfo {year} {1990})}\BibitemShut {NoStop}%
\bibitem [{\citenamefont {{Kitaura}}\ \emph {et~al.}(2016)\citenamefont
  {{Kitaura}}, \citenamefont {{Chuang}}, \citenamefont {{Liang}}, \citenamefont
  {{Zhao}}, \citenamefont {{Tao}}, \citenamefont {{Rodr{\'{\i}}guez-Torres}},
  \citenamefont {{Eisenstein}}, \citenamefont {{Gil-Mar{\'{\i}}n}},
  \citenamefont {{Kneib}}, \citenamefont {{McBride}}, \citenamefont
  {{Percival}}, \citenamefont {{Ross}}, \citenamefont {{S{\'a}nchez}},
  \citenamefont {{Tinker}}, \citenamefont {{Tojeiro}}, \citenamefont
  {{Vargas-Magana}},\ and\ \citenamefont {{Zhao}}}]{Kitauraetal2016b}%
  \BibitemOpen
  \bibfield  {author} {\bibinfo {author} {\bibfnamefont {F.-S.}\ \bibnamefont
  {{Kitaura}}}, \bibinfo {author} {\bibfnamefont {C.-H.}\ \bibnamefont
  {{Chuang}}}, \bibinfo {author} {\bibfnamefont {Y.}~\bibnamefont {{Liang}}},
  \bibinfo {author} {\bibfnamefont {C.}~\bibnamefont {{Zhao}}}, \bibinfo
  {author} {\bibfnamefont {C.}~\bibnamefont {{Tao}}}, \bibinfo {author}
  {\bibfnamefont {S.}~\bibnamefont {{Rodr{\'{\i}}guez-Torres}}}, \bibinfo
  {author} {\bibfnamefont {D.~J.}\ \bibnamefont {{Eisenstein}}}, \bibinfo
  {author} {\bibfnamefont {H.}~\bibnamefont {{Gil-Mar{\'{\i}}n}}}, \bibinfo
  {author} {\bibfnamefont {J.-P.}\ \bibnamefont {{Kneib}}}, \bibinfo {author}
  {\bibfnamefont {C.}~\bibnamefont {{McBride}}}, \bibinfo {author}
  {\bibfnamefont {W.~J.}\ \bibnamefont {{Percival}}}, \bibinfo {author}
  {\bibfnamefont {A.~J.}\ \bibnamefont {{Ross}}}, \bibinfo {author}
  {\bibfnamefont {A.~G.}\ \bibnamefont {{S{\'a}nchez}}}, \bibinfo {author}
  {\bibfnamefont {J.}~\bibnamefont {{Tinker}}}, \bibinfo {author}
  {\bibfnamefont {R.}~\bibnamefont {{Tojeiro}}}, \bibinfo {author}
  {\bibfnamefont {M.}~\bibnamefont {{Vargas-Magana}}}, \ and\ \bibinfo {author}
  {\bibfnamefont {G.-B.}\ \bibnamefont {{Zhao}}},\ }\href {\doibase
  10.1103/PhysRevLett.116.171301} {\bibfield  {journal} {\bibinfo  {journal}
  {Physical Review Letters}\ }\textbf {\bibinfo {volume} {116}},\ \bibinfo
  {eid} {171301} (\bibinfo {year} {2016})}\BibitemShut {NoStop}%
\bibitem [{\citenamefont {{Liang}}\ \emph {et~al.}(2016)\citenamefont
  {{Liang}}, \citenamefont {{Zhao}}, \citenamefont {{Chuang}}, \citenamefont
  {{Kitaura}},\ and\ \citenamefont {{Tao}}}]{Liang2016}%
  \BibitemOpen
  \bibfield  {author} {\bibinfo {author} {\bibfnamefont {Y.}~\bibnamefont
  {{Liang}}}, \bibinfo {author} {\bibfnamefont {C.}~\bibnamefont {{Zhao}}},
  \bibinfo {author} {\bibfnamefont {C.-H.}\ \bibnamefont {{Chuang}}}, \bibinfo
  {author} {\bibfnamefont {F.-S.}\ \bibnamefont {{Kitaura}}}, \ and\ \bibinfo
  {author} {\bibfnamefont {C.}~\bibnamefont {{Tao}}},\ }\href {\doibase
  10.1093/mnras/stw884} {\bibfield  {journal} {\bibinfo  {journal} {\mnras}\
  }\textbf {\bibinfo {volume} {459}},\ \bibinfo {pages} {4020} (\bibinfo {year}
  {2016})},\ \Eprint {http://arxiv.org/abs/1511.04391} {arXiv:1511.04391}
  \BibitemShut {NoStop}%
\bibitem [{\citenamefont {{Sheth}}\ and\ \citenamefont {{van de
  Weygaert}}(2004)}]{SW04}%
  \BibitemOpen
  \bibfield  {author} {\bibinfo {author} {\bibfnamefont {R.~K.}\ \bibnamefont
  {{Sheth}}}\ and\ \bibinfo {author} {\bibfnamefont {R.}~\bibnamefont {{van de
  Weygaert}}},\ }\href {\doibase 10.1111/j.1365-2966.2004.07661.x} {\bibfield
  {journal} {\bibinfo  {journal} {\mnras}\ }\textbf {\bibinfo {volume} {350}},\
  \bibinfo {pages} {517} (\bibinfo {year} {2004})},\ \Eprint
  {http://arxiv.org/abs/astro-ph/0311260} {astro-ph/0311260} \BibitemShut
  {NoStop}%
\bibitem [{\citenamefont {{Hamaus}}\ \emph {et~al.}(2015)\citenamefont
  {{Hamaus}}, \citenamefont {{Sutter}}, \citenamefont {{Lavaux}},\ and\
  \citenamefont {{Wandelt}}}]{Hamaus2015}%
  \BibitemOpen
  \bibfield  {author} {\bibinfo {author} {\bibfnamefont {N.}~\bibnamefont
  {{Hamaus}}}, \bibinfo {author} {\bibfnamefont {P.~M.}\ \bibnamefont
  {{Sutter}}}, \bibinfo {author} {\bibfnamefont {G.}~\bibnamefont {{Lavaux}}},
  \ and\ \bibinfo {author} {\bibfnamefont {B.~D.}\ \bibnamefont {{Wandelt}}},\
  }\href {\doibase 10.1088/1475-7516/2015/11/036} {\bibfield  {journal}
  {\bibinfo  {journal} {\jcap}\ }\textbf {\bibinfo {volume} {11}},\ \bibinfo
  {eid} {036} (\bibinfo {year} {2015})},\ \Eprint
  {http://arxiv.org/abs/1507.04363} {arXiv:1507.04363} \BibitemShut {NoStop}%
\bibitem [{\citenamefont {{Wojtak}}\ \emph {et~al.}(2016)\citenamefont
  {{Wojtak}}, \citenamefont {{Powell}},\ and\ \citenamefont
  {{Abel}}}]{Wojtak2016}%
  \BibitemOpen
  \bibfield  {author} {\bibinfo {author} {\bibfnamefont {R.}~\bibnamefont
  {{Wojtak}}}, \bibinfo {author} {\bibfnamefont {D.}~\bibnamefont {{Powell}}},
  \ and\ \bibinfo {author} {\bibfnamefont {T.}~\bibnamefont {{Abel}}},\ }\href
  {\doibase 10.1093/mnras/stw615} {\bibfield  {journal} {\bibinfo  {journal}
  {\mnras}\ }\textbf {\bibinfo {volume} {458}},\ \bibinfo {pages} {4431}
  (\bibinfo {year} {2016})},\ \Eprint {http://arxiv.org/abs/1602.08541}
  {arXiv:1602.08541} \BibitemShut {NoStop}%
\bibitem [{\citenamefont {{Kaiser}}(1987)}]{Kaiser1987}%
  \BibitemOpen
  \bibfield  {author} {\bibinfo {author} {\bibfnamefont {N.}~\bibnamefont
  {{Kaiser}}},\ }\href {\doibase 10.1093/mnras/227.1.1} {\bibfield  {journal}
  {\bibinfo  {journal} {\mnras}\ }\textbf {\bibinfo {volume} {227}},\ \bibinfo
  {pages} {1} (\bibinfo {year} {1987})}\BibitemShut {NoStop}%
\bibitem [{\citenamefont {{Hamilton}}(1992)}]{Hamilton1992}%
  \BibitemOpen
  \bibfield  {author} {\bibinfo {author} {\bibfnamefont {A.~J.~S.}\
  \bibnamefont {{Hamilton}}},\ }\href {\doibase 10.1086/186264} {\bibfield
  {journal} {\bibinfo  {journal} {\apjl}\ }\textbf {\bibinfo {volume} {385}},\
  \bibinfo {pages} {L5} (\bibinfo {year} {1992})}\BibitemShut {NoStop}%
\bibitem [{\citenamefont {{Cole}}\ \emph {et~al.}(1994)\citenamefont {{Cole}},
  \citenamefont {{Fisher}},\ and\ \citenamefont {{Weinberg}}}]{Cole1994}%
  \BibitemOpen
  \bibfield  {author} {\bibinfo {author} {\bibfnamefont {S.}~\bibnamefont
  {{Cole}}}, \bibinfo {author} {\bibfnamefont {K.~B.}\ \bibnamefont
  {{Fisher}}}, \ and\ \bibinfo {author} {\bibfnamefont {D.~H.}\ \bibnamefont
  {{Weinberg}}},\ }\href {\doibase 10.1093/mnras/267.3.785} {\bibfield
  {journal} {\bibinfo  {journal} {\mnras}\ }\textbf {\bibinfo {volume} {267}},\
  \bibinfo {pages} {785} (\bibinfo {year} {1994})},\ \Eprint
  {http://arxiv.org/abs/astro-ph/9308003} {astro-ph/9308003} \BibitemShut
  {NoStop}%
\bibitem [{\citenamefont {{Cole}}\ \emph {et~al.}(1995)\citenamefont {{Cole}},
  \citenamefont {{Fisher}},\ and\ \citenamefont {{Weinberg}}}]{Cole1995}%
  \BibitemOpen
  \bibfield  {author} {\bibinfo {author} {\bibfnamefont {S.}~\bibnamefont
  {{Cole}}}, \bibinfo {author} {\bibfnamefont {K.~B.}\ \bibnamefont
  {{Fisher}}}, \ and\ \bibinfo {author} {\bibfnamefont {D.~H.}\ \bibnamefont
  {{Weinberg}}},\ }\href {\doibase 10.1093/mnras/275.2.515} {\bibfield
  {journal} {\bibinfo  {journal} {\mnras}\ }\textbf {\bibinfo {volume} {275}},\
  \bibinfo {pages} {515} (\bibinfo {year} {1995})},\ \Eprint
  {http://arxiv.org/abs/astro-ph/9412062} {astro-ph/9412062} \BibitemShut
  {NoStop}%
\bibitem [{\citenamefont {{Peacock et al.}}(2001)}]{Peacocketal2001}%
  \BibitemOpen
  \bibfield  {author} {\bibinfo {author} {\bibfnamefont {J.~A.}\ \bibnamefont
  {{Peacock et al.}}},\ }\href@noop {} {\bibfield  {journal} {\bibinfo
  {journal} {\nat}\ }\textbf {\bibinfo {volume} {410}},\ \bibinfo {pages} {169}
  (\bibinfo {year} {2001})},\ \Eprint {http://arxiv.org/abs/astro-ph/0103143}
  {astro-ph/0103143} \BibitemShut {NoStop}%
\bibitem [{\citenamefont {{Scoccimarro}}(2004)}]{Scoccimarro2004}%
  \BibitemOpen
  \bibfield  {author} {\bibinfo {author} {\bibfnamefont {R.}~\bibnamefont
  {{Scoccimarro}}},\ }\href {\doibase 10.1103/PhysRevD.70.083007} {\bibfield
  {journal} {\bibinfo  {journal} {\prd}\ }\textbf {\bibinfo {volume} {70}},\
  \bibinfo {eid} {083007} (\bibinfo {year} {2004})},\ \Eprint
  {http://arxiv.org/abs/astro-ph/0407214} {astro-ph/0407214} \BibitemShut
  {NoStop}%
\bibitem [{\citenamefont {{Guzzo}}\ \emph {et~al.}(2008)\citenamefont
  {{Guzzo}}, \citenamefont {{Pierleoni}}, \citenamefont {{Meneux}},
  \citenamefont {{Branchini}}, \citenamefont {{Le F{\`e}vre}}, \citenamefont
  {{Marinoni}}, \citenamefont {{Garilli}}, \citenamefont {{Blaizot}},\ and\
  \citenamefont {{et al}}}]{Guzzo2008}%
  \BibitemOpen
  \bibfield  {author} {\bibinfo {author} {\bibfnamefont {L.}~\bibnamefont
  {{Guzzo}}}, \bibinfo {author} {\bibfnamefont {M.}~\bibnamefont
  {{Pierleoni}}}, \bibinfo {author} {\bibfnamefont {B.}~\bibnamefont
  {{Meneux}}}, \bibinfo {author} {\bibfnamefont {E.}~\bibnamefont
  {{Branchini}}}, \bibinfo {author} {\bibfnamefont {O.}~\bibnamefont {{Le
  F{\`e}vre}}}, \bibinfo {author} {\bibfnamefont {C.}~\bibnamefont
  {{Marinoni}}}, \bibinfo {author} {\bibfnamefont {B.}~\bibnamefont
  {{Garilli}}}, \bibinfo {author} {\bibfnamefont {J.}~\bibnamefont
  {{Blaizot}}}, \ and\ \bibinfo {author} {\bibnamefont {{et al}}},\ }\href
  {\doibase 10.1038/nature06555} {\bibfield  {journal} {\bibinfo  {journal}
  {\nat}\ }\textbf {\bibinfo {volume} {451}},\ \bibinfo {pages} {541} (\bibinfo
  {year} {2008})},\ \Eprint {http://arxiv.org/abs/0802.1944} {arXiv:0802.1944}
  \BibitemShut {NoStop}%
\bibitem [{\citenamefont {{Matsubara}}(2008)}]{Matsubara2008}%
  \BibitemOpen
  \bibfield  {author} {\bibinfo {author} {\bibfnamefont {T.}~\bibnamefont
  {{Matsubara}}},\ }\href {\doibase 10.1103/PhysRevD.78.083519} {\bibfield
  {journal} {\bibinfo  {journal} {\prd}\ }\textbf {\bibinfo {volume} {78}},\
  \bibinfo {eid} {083519} (\bibinfo {year} {2008})},\ \Eprint
  {http://arxiv.org/abs/0807.1733} {arXiv:0807.1733} \BibitemShut {NoStop}%
\bibitem [{\citenamefont {{Taruya}}\ \emph {et~al.}(2009)\citenamefont
  {{Taruya}}, \citenamefont {{Nishimichi}}, \citenamefont {{Saito}},\ and\
  \citenamefont {{Hiramatsu}}}]{Taruya2009}%
  \BibitemOpen
  \bibfield  {author} {\bibinfo {author} {\bibfnamefont {A.}~\bibnamefont
  {{Taruya}}}, \bibinfo {author} {\bibfnamefont {T.}~\bibnamefont
  {{Nishimichi}}}, \bibinfo {author} {\bibfnamefont {S.}~\bibnamefont
  {{Saito}}}, \ and\ \bibinfo {author} {\bibfnamefont {T.}~\bibnamefont
  {{Hiramatsu}}},\ }\href {\doibase 10.1103/PhysRevD.80.123503} {\bibfield
  {journal} {\bibinfo  {journal} {\prd}\ }\textbf {\bibinfo {volume} {80}},\
  \bibinfo {eid} {123503} (\bibinfo {year} {2009})},\ \Eprint
  {http://arxiv.org/abs/0906.0507} {arXiv:0906.0507 [astro-ph.CO]} \BibitemShut
  {NoStop}%
\bibitem [{\citenamefont {{Percival}}\ and\ \citenamefont
  {{White}}(2009)}]{Percival2009}%
  \BibitemOpen
  \bibfield  {author} {\bibinfo {author} {\bibfnamefont {W.~J.}\ \bibnamefont
  {{Percival}}}\ and\ \bibinfo {author} {\bibfnamefont {M.}~\bibnamefont
  {{White}}},\ }\href {\doibase 10.1111/j.1365-2966.2008.14211.x} {\bibfield
  {journal} {\bibinfo  {journal} {\mnras}\ }\textbf {\bibinfo {volume} {393}},\
  \bibinfo {pages} {297} (\bibinfo {year} {2009})},\ \Eprint
  {http://arxiv.org/abs/0808.0003} {arXiv:0808.0003} \BibitemShut {NoStop}%
\bibitem [{\citenamefont {{Taruya}}\ \emph {et~al.}(2010)\citenamefont
  {{Taruya}}, \citenamefont {{Nishimichi}},\ and\ \citenamefont
  {{Saito}}}]{Taruya2010}%
  \BibitemOpen
  \bibfield  {author} {\bibinfo {author} {\bibfnamefont {A.}~\bibnamefont
  {{Taruya}}}, \bibinfo {author} {\bibfnamefont {T.}~\bibnamefont
  {{Nishimichi}}}, \ and\ \bibinfo {author} {\bibfnamefont {S.}~\bibnamefont
  {{Saito}}},\ }\href {\doibase 10.1103/PhysRevD.82.063522} {\bibfield
  {journal} {\bibinfo  {journal} {\prd}\ }\textbf {\bibinfo {volume} {82}},\
  \bibinfo {eid} {063522} (\bibinfo {year} {2010})},\ \Eprint
  {http://arxiv.org/abs/1006.0699} {arXiv:1006.0699} \BibitemShut {NoStop}%
\bibitem [{\citenamefont {{Seljak}}\ and\ \citenamefont
  {{McDonald}}(2011)}]{Seljak2011}%
  \BibitemOpen
  \bibfield  {author} {\bibinfo {author} {\bibfnamefont {U.}~\bibnamefont
  {{Seljak}}}\ and\ \bibinfo {author} {\bibfnamefont {P.}~\bibnamefont
  {{McDonald}}},\ }\href {\doibase 10.1088/1475-7516/2011/11/039} {\bibfield
  {journal} {\bibinfo  {journal} {\jcap}\ }\textbf {\bibinfo {volume} {11}},\
  \bibinfo {eid} {039} (\bibinfo {year} {2011})},\ \Eprint
  {http://arxiv.org/abs/1109.1888} {arXiv:1109.1888} \BibitemShut {NoStop}%
\bibitem [{\citenamefont {{Jennings}}\ \emph {et~al.}(2011)\citenamefont
  {{Jennings}}, \citenamefont {{Baugh}},\ and\ \citenamefont
  {{Pascoli}}}]{Jennings2011}%
  \BibitemOpen
  \bibfield  {author} {\bibinfo {author} {\bibfnamefont {E.}~\bibnamefont
  {{Jennings}}}, \bibinfo {author} {\bibfnamefont {C.~M.}\ \bibnamefont
  {{Baugh}}}, \ and\ \bibinfo {author} {\bibfnamefont {S.}~\bibnamefont
  {{Pascoli}}},\ }\href {\doibase 10.1111/j.1365-2966.2010.17581.x} {\bibfield
  {journal} {\bibinfo  {journal} {\mnras}\ }\textbf {\bibinfo {volume} {410}},\
  \bibinfo {pages} {2081} (\bibinfo {year} {2011})},\ \Eprint
  {http://arxiv.org/abs/1003.4282} {arXiv:1003.4282 [astro-ph.CO]} \BibitemShut
  {NoStop}%
\bibitem [{\citenamefont {{Gil-Mar{\'{\i}}n}}\ \emph
  {et~al.}(2012)\citenamefont {{Gil-Mar{\'{\i}}n}}, \citenamefont {{Wagner}},
  \citenamefont {{Verde}}, \citenamefont {{Porciani}},\ and\ \citenamefont
  {{Jimenez}}}]{Gil-Marin2012}%
  \BibitemOpen
  \bibfield  {author} {\bibinfo {author} {\bibfnamefont {H.}~\bibnamefont
  {{Gil-Mar{\'{\i}}n}}}, \bibinfo {author} {\bibfnamefont {C.}~\bibnamefont
  {{Wagner}}}, \bibinfo {author} {\bibfnamefont {L.}~\bibnamefont {{Verde}}},
  \bibinfo {author} {\bibfnamefont {C.}~\bibnamefont {{Porciani}}}, \ and\
  \bibinfo {author} {\bibfnamefont {R.}~\bibnamefont {{Jimenez}}},\ }\href
  {\doibase 10.1088/1475-7516/2012/11/029} {\bibfield  {journal} {\bibinfo
  {journal} {\jcap}\ }\textbf {\bibinfo {volume} {11}},\ \bibinfo {eid} {029}
  (\bibinfo {year} {2012})},\ \Eprint {http://arxiv.org/abs/1209.3771}
  {arXiv:1209.3771 [astro-ph.CO]} \BibitemShut {NoStop}%
\bibitem [{\citenamefont {{Beutler}}\ \emph {et~al.}(2012)\citenamefont
  {{Beutler}}, \citenamefont {{Blake}}, \citenamefont {{Colless}},
  \citenamefont {{Jones}}, \citenamefont {{Staveley-Smith}}, \citenamefont
  {{Poole}}, \citenamefont {{Campbell}}, \citenamefont {{Parker}},
  \citenamefont {{Saunders}},\ and\ \citenamefont {{Watson}}}]{Beutler2012}%
  \BibitemOpen
  \bibfield  {author} {\bibinfo {author} {\bibfnamefont {F.}~\bibnamefont
  {{Beutler}}}, \bibinfo {author} {\bibfnamefont {C.}~\bibnamefont {{Blake}}},
  \bibinfo {author} {\bibfnamefont {M.}~\bibnamefont {{Colless}}}, \bibinfo
  {author} {\bibfnamefont {D.~H.}\ \bibnamefont {{Jones}}}, \bibinfo {author}
  {\bibfnamefont {L.}~\bibnamefont {{Staveley-Smith}}}, \bibinfo {author}
  {\bibfnamefont {G.~B.}\ \bibnamefont {{Poole}}}, \bibinfo {author}
  {\bibfnamefont {L.}~\bibnamefont {{Campbell}}}, \bibinfo {author}
  {\bibfnamefont {Q.}~\bibnamefont {{Parker}}}, \bibinfo {author}
  {\bibfnamefont {W.}~\bibnamefont {{Saunders}}}, \ and\ \bibinfo {author}
  {\bibfnamefont {F.}~\bibnamefont {{Watson}}},\ }\href {\doibase
  10.1111/j.1365-2966.2012.21136.x} {\bibfield  {journal} {\bibinfo  {journal}
  {\mnras}\ }\textbf {\bibinfo {volume} {423}},\ \bibinfo {pages} {3430}
  (\bibinfo {year} {2012})},\ \Eprint {http://arxiv.org/abs/1204.4725}
  {arXiv:1204.4725} \BibitemShut {NoStop}%
\bibitem [{\citenamefont {{Reid et al.}}(2012)}]{Reidetal2012}%
  \BibitemOpen
  \bibfield  {author} {\bibinfo {author} {\bibfnamefont {B.~A.}\ \bibnamefont
  {{Reid et al.}}},\ }\href {\doibase 10.1111/j.1365-2966.2012.21779.x}
  {\bibfield  {journal} {\bibinfo  {journal} {\mnras}\ }\textbf {\bibinfo
  {volume} {426}},\ \bibinfo {pages} {2719} (\bibinfo {year} {2012})},\ \Eprint
  {http://arxiv.org/abs/1203.6641} {arXiv:1203.6641} \BibitemShut {NoStop}%
\bibitem [{\citenamefont {{Blake et al.}}(2012)}]{Blakeetal2012}%
  \BibitemOpen
  \bibfield  {author} {\bibinfo {author} {\bibfnamefont {C.}~\bibnamefont
  {{Blake et al.}}},\ }\href {\doibase 10.1111/j.1365-2966.2012.21473.x}
  {\bibfield  {journal} {\bibinfo  {journal} {\mnras}\ }\textbf {\bibinfo
  {volume} {425}},\ \bibinfo {pages} {405} (\bibinfo {year} {2012})},\ \Eprint
  {http://arxiv.org/abs/1204.3674} {arXiv:1204.3674} \BibitemShut {NoStop}%
\bibitem [{\citenamefont {{Okumura}}\ \emph {et~al.}(2012)\citenamefont
  {{Okumura}}, \citenamefont {{Seljak}}, \citenamefont {{McDonald}},\ and\
  \citenamefont {{Desjacques}}}]{Okumura2012}%
  \BibitemOpen
  \bibfield  {author} {\bibinfo {author} {\bibfnamefont {T.}~\bibnamefont
  {{Okumura}}}, \bibinfo {author} {\bibfnamefont {U.}~\bibnamefont {{Seljak}}},
  \bibinfo {author} {\bibfnamefont {P.}~\bibnamefont {{McDonald}}}, \ and\
  \bibinfo {author} {\bibfnamefont {V.}~\bibnamefont {{Desjacques}}},\ }\href
  {\doibase 10.1088/1475-7516/2012/02/010} {\bibfield  {journal} {\bibinfo
  {journal} {\jcap}\ }\textbf {\bibinfo {volume} {2}},\ \bibinfo {eid} {010}
  (\bibinfo {year} {2012})},\ \Eprint {http://arxiv.org/abs/1109.1609}
  {arXiv:1109.1609} \BibitemShut {NoStop}%
\bibitem [{\citenamefont {{de la Torre}}\ and\ \citenamefont
  {{Guzzo}}(2012)}]{delaTorre2012}%
  \BibitemOpen
  \bibfield  {author} {\bibinfo {author} {\bibfnamefont {S.}~\bibnamefont {{de
  la Torre}}}\ and\ \bibinfo {author} {\bibfnamefont {L.}~\bibnamefont
  {{Guzzo}}},\ }\href {\doibase 10.1111/j.1365-2966.2012.21824.x} {\bibfield
  {journal} {\bibinfo  {journal} {\mnras}\ }\textbf {\bibinfo {volume} {427}},\
  \bibinfo {pages} {327} (\bibinfo {year} {2012})},\ \Eprint
  {http://arxiv.org/abs/1202.5559} {arXiv:1202.5559} \BibitemShut {NoStop}%
\bibitem [{\citenamefont {{Valageas}}\ \emph {et~al.}(2013)\citenamefont
  {{Valageas}}, \citenamefont {{Nishimichi}},\ and\ \citenamefont
  {{Taruya}}}]{Valageas2013}%
  \BibitemOpen
  \bibfield  {author} {\bibinfo {author} {\bibfnamefont {P.}~\bibnamefont
  {{Valageas}}}, \bibinfo {author} {\bibfnamefont {T.}~\bibnamefont
  {{Nishimichi}}}, \ and\ \bibinfo {author} {\bibfnamefont {A.}~\bibnamefont
  {{Taruya}}},\ }\href {\doibase 10.1103/PhysRevD.87.083522} {\bibfield
  {journal} {\bibinfo  {journal} {\prd}\ }\textbf {\bibinfo {volume} {87}},\
  \bibinfo {eid} {083522} (\bibinfo {year} {2013})},\ \Eprint
  {http://arxiv.org/abs/1302.4533} {arXiv:1302.4533 [astro-ph.CO]} \BibitemShut
  {NoStop}%
\bibitem [{\citenamefont {{Chuang}}\ and\ \citenamefont
  {{Wang}}(2013{\natexlab{a}})}]{Chuang2013}%
  \BibitemOpen
  \bibfield  {author} {\bibinfo {author} {\bibfnamefont {C.-H.}\ \bibnamefont
  {{Chuang}}}\ and\ \bibinfo {author} {\bibfnamefont {Y.}~\bibnamefont
  {{Wang}}},\ }\href {\doibase 10.1093/mnras/stt357} {\bibfield  {journal}
  {\bibinfo  {journal} {\mnras}\ }\textbf {\bibinfo {volume} {431}},\ \bibinfo
  {pages} {2634} (\bibinfo {year} {2013}{\natexlab{a}})},\ \Eprint
  {http://arxiv.org/abs/1205.5573} {arXiv:1205.5573} \BibitemShut {NoStop}%
\bibitem [{\citenamefont {{Chuang}}\ and\ \citenamefont
  {{Wang}}(2013{\natexlab{b}})}]{Chuang2013b}%
  \BibitemOpen
  \bibfield  {author} {\bibinfo {author} {\bibfnamefont {C.-H.}\ \bibnamefont
  {{Chuang}}}\ and\ \bibinfo {author} {\bibfnamefont {Y.}~\bibnamefont
  {{Wang}}},\ }\href {\doibase 10.1093/mnras/stt1290} {\bibfield  {journal}
  {\bibinfo  {journal} {\mnras}\ }\textbf {\bibinfo {volume} {435}},\ \bibinfo
  {pages} {255} (\bibinfo {year} {2013}{\natexlab{b}})},\ \Eprint
  {http://arxiv.org/abs/1209.0210} {arXiv:1209.0210} \BibitemShut {NoStop}%
\bibitem [{\citenamefont {{Chuang}}\ \emph {et~al.}(2013)\citenamefont
  {{Chuang}}, \citenamefont {{Prada}}, \citenamefont {{Cuesta}}, \citenamefont
  {{Eisenstein}}, \citenamefont {{Kazin}}, \citenamefont {{Padmanabhan}},
  \citenamefont {{S{\'a}nchez}}, \citenamefont {{Xu}},\ and\ \citenamefont {{et
  al}}}]{Chuang2013c}%
  \BibitemOpen
  \bibfield  {author} {\bibinfo {author} {\bibfnamefont {C.-H.}\ \bibnamefont
  {{Chuang}}}, \bibinfo {author} {\bibfnamefont {F.}~\bibnamefont {{Prada}}},
  \bibinfo {author} {\bibfnamefont {A.~J.}\ \bibnamefont {{Cuesta}}}, \bibinfo
  {author} {\bibfnamefont {D.~J.}\ \bibnamefont {{Eisenstein}}}, \bibinfo
  {author} {\bibfnamefont {E.}~\bibnamefont {{Kazin}}}, \bibinfo {author}
  {\bibfnamefont {N.}~\bibnamefont {{Padmanabhan}}}, \bibinfo {author}
  {\bibfnamefont {A.~G.}\ \bibnamefont {{S{\'a}nchez}}}, \bibinfo {author}
  {\bibfnamefont {X.}~\bibnamefont {{Xu}}}, \ and\ \bibinfo {author}
  {\bibnamefont {{et al}}},\ }\href {\doibase 10.1093/mnras/stt988} {\bibfield
  {journal} {\bibinfo  {journal} {\mnras}\ }\textbf {\bibinfo {volume} {433}},\
  \bibinfo {pages} {3559} (\bibinfo {year} {2013})},\ \Eprint
  {http://arxiv.org/abs/1303.4486} {arXiv:1303.4486} \BibitemShut {NoStop}%
\bibitem [{\citenamefont {{Chuang}}\ \emph {et~al.}(2016)\citenamefont
  {{Chuang}}, \citenamefont {{Prada}}, \citenamefont {{Pellejero-Ibanez}},
  \citenamefont {{Beutler}}, \citenamefont {{Cuesta}}, \citenamefont
  {{Eisenstein}}, \citenamefont {{Escoffier}}, \citenamefont {{Ho}},\ and\
  \citenamefont {{et al}}}]{Chuang2016}%
  \BibitemOpen
  \bibfield  {author} {\bibinfo {author} {\bibfnamefont {C.-H.}\ \bibnamefont
  {{Chuang}}}, \bibinfo {author} {\bibfnamefont {F.}~\bibnamefont {{Prada}}},
  \bibinfo {author} {\bibfnamefont {M.}~\bibnamefont {{Pellejero-Ibanez}}},
  \bibinfo {author} {\bibfnamefont {F.}~\bibnamefont {{Beutler}}}, \bibinfo
  {author} {\bibfnamefont {A.~J.}\ \bibnamefont {{Cuesta}}}, \bibinfo {author}
  {\bibfnamefont {D.~J.}\ \bibnamefont {{Eisenstein}}}, \bibinfo {author}
  {\bibfnamefont {S.}~\bibnamefont {{Escoffier}}}, \bibinfo {author}
  {\bibfnamefont {S.}~\bibnamefont {{Ho}}}, \ and\ \bibinfo {author}
  {\bibnamefont {{et al}}},\ }\href@noop {} {\bibfield  {journal} {\bibinfo
  {journal} {\mnras}\ }\textbf {\bibinfo {volume} {461}},\ \bibinfo {pages}
  {3781} (\bibinfo {year} {2016})},\ \Eprint {http://arxiv.org/abs/1312.4889}
  {arXiv:1312.4889} \BibitemShut {NoStop}%
\bibitem [{\citenamefont {{de la Torre et al.}}(2013)}]{delaTorreetal2013}%
  \BibitemOpen
  \bibfield  {author} {\bibinfo {author} {\bibfnamefont {S.}~\bibnamefont {{de
  la Torre et al.}}},\ }\href {\doibase 10.1051/0004-6361/201321463} {\bibfield
   {journal} {\bibinfo  {journal} {\aap}\ }\textbf {\bibinfo {volume} {557}},\
  \bibinfo {eid} {A54} (\bibinfo {year} {2013})},\ \Eprint
  {http://arxiv.org/abs/1303.2622} {arXiv:1303.2622} \BibitemShut {NoStop}%
\bibitem [{\citenamefont {{Samushia et al.}}(2014)}]{Samushiaetal2014}%
  \BibitemOpen
  \bibfield  {author} {\bibinfo {author} {\bibfnamefont {L.}~\bibnamefont
  {{Samushia et al.}}},\ }\href {\doibase 10.1093/mnras/stu197} {\bibfield
  {journal} {\bibinfo  {journal} {\mnras}\ }\textbf {\bibinfo {volume} {439}},\
  \bibinfo {pages} {3504} (\bibinfo {year} {2014})},\ \Eprint
  {http://arxiv.org/abs/1312.4899} {arXiv:1312.4899} \BibitemShut {NoStop}%
\bibitem [{\citenamefont {{Howlett}}\ \emph {et~al.}(2015)\citenamefont
  {{Howlett}}, \citenamefont {{Ross}}, \citenamefont {{Samushia}},
  \citenamefont {{Percival}},\ and\ \citenamefont {{Manera}}}]{Howlett2015}%
  \BibitemOpen
  \bibfield  {author} {\bibinfo {author} {\bibfnamefont {C.}~\bibnamefont
  {{Howlett}}}, \bibinfo {author} {\bibfnamefont {A.~J.}\ \bibnamefont
  {{Ross}}}, \bibinfo {author} {\bibfnamefont {L.}~\bibnamefont {{Samushia}}},
  \bibinfo {author} {\bibfnamefont {W.~J.}\ \bibnamefont {{Percival}}}, \ and\
  \bibinfo {author} {\bibfnamefont {M.}~\bibnamefont {{Manera}}},\ }\href
  {\doibase 10.1093/mnras/stu2693} {\bibfield  {journal} {\bibinfo  {journal}
  {\mnras}\ }\textbf {\bibinfo {volume} {449}},\ \bibinfo {pages} {848}
  (\bibinfo {year} {2015})},\ \Eprint {http://arxiv.org/abs/1409.3238}
  {arXiv:1409.3238} \BibitemShut {NoStop}%
\bibitem [{\citenamefont {{Reid}}\ \emph {et~al.}(2014)\citenamefont {{Reid}},
  \citenamefont {{Seo}}, \citenamefont {{Leauthaud}}, \citenamefont
  {{Tinker}},\ and\ \citenamefont {{White}}}]{Reid2014}%
  \BibitemOpen
  \bibfield  {author} {\bibinfo {author} {\bibfnamefont {B.~A.}\ \bibnamefont
  {{Reid}}}, \bibinfo {author} {\bibfnamefont {H.-J.}\ \bibnamefont {{Seo}}},
  \bibinfo {author} {\bibfnamefont {A.}~\bibnamefont {{Leauthaud}}}, \bibinfo
  {author} {\bibfnamefont {J.~L.}\ \bibnamefont {{Tinker}}}, \ and\ \bibinfo
  {author} {\bibfnamefont {M.}~\bibnamefont {{White}}},\ }\href {\doibase
  10.1093/mnras/stu1391} {\bibfield  {journal} {\bibinfo  {journal} {\mnras}\
  }\textbf {\bibinfo {volume} {444}},\ \bibinfo {pages} {476} (\bibinfo {year}
  {2014})},\ \Eprint {http://arxiv.org/abs/1404.3742} {arXiv:1404.3742}
  \BibitemShut {NoStop}%
\bibitem [{\citenamefont {{Okumura}}\ \emph {et~al.}(2015)\citenamefont
  {{Okumura}}, \citenamefont {{Hand}}, \citenamefont {{Seljak}}, \citenamefont
  {{Vlah}},\ and\ \citenamefont {{Desjacques}}}]{Okumura2015a}%
  \BibitemOpen
  \bibfield  {author} {\bibinfo {author} {\bibfnamefont {T.}~\bibnamefont
  {{Okumura}}}, \bibinfo {author} {\bibfnamefont {N.}~\bibnamefont {{Hand}}},
  \bibinfo {author} {\bibfnamefont {U.}~\bibnamefont {{Seljak}}}, \bibinfo
  {author} {\bibfnamefont {Z.}~\bibnamefont {{Vlah}}}, \ and\ \bibinfo {author}
  {\bibfnamefont {V.}~\bibnamefont {{Desjacques}}},\ }\href {\doibase
  10.1103/PhysRevD.92.103516} {\bibfield  {journal} {\bibinfo  {journal}
  {\prd}\ }\textbf {\bibinfo {volume} {92}},\ \bibinfo {eid} {103516} (\bibinfo
  {year} {2015})},\ \Eprint {http://arxiv.org/abs/1506.05814}
  {arXiv:1506.05814} \BibitemShut {NoStop}%
\bibitem [{\citenamefont {Okumura}\ \emph {et~al.}(2016)\citenamefont {Okumura}
  \emph {et~al.}}]{Okumura:2015lvp}%
  \BibitemOpen
  \bibfield  {author} {\bibinfo {author} {\bibfnamefont {T.}~\bibnamefont
  {Okumura}} \emph {et~al.},\ }\href {\doibase 10.1093/pasj/psw029} {\bibfield
  {journal} {\bibinfo  {journal} {Publ. Astron. Soc. Jap.}\ }\textbf {\bibinfo
  {volume} {68}},\ \bibinfo {pages} {24} (\bibinfo {year} {2016})},\ \Eprint
  {http://arxiv.org/abs/1511.08083} {arXiv:1511.08083 [astro-ph.CO]}
  \BibitemShut {NoStop}%
%%CITATION = ARXIV:1511.08083;%%
\bibitem [{\citenamefont {{Alam}}\ \emph
  {et~al.}(2015{\natexlab{a}})\citenamefont {{Alam}}, \citenamefont {{Ho}},
  \citenamefont {{Vargas-Maga{\~n}a}},\ and\ \citenamefont
  {{Schneider}}}]{Alam2015}%
  \BibitemOpen
  \bibfield  {author} {\bibinfo {author} {\bibfnamefont {S.}~\bibnamefont
  {{Alam}}}, \bibinfo {author} {\bibfnamefont {S.}~\bibnamefont {{Ho}}},
  \bibinfo {author} {\bibfnamefont {M.}~\bibnamefont {{Vargas-Maga{\~n}a}}}, \
  and\ \bibinfo {author} {\bibfnamefont {D.~P.}\ \bibnamefont {{Schneider}}},\
  }\href {\doibase 10.1093/mnras/stv1737} {\bibfield  {journal} {\bibinfo
  {journal} {\mnras}\ }\textbf {\bibinfo {volume} {453}},\ \bibinfo {pages}
  {1754} (\bibinfo {year} {2015}{\natexlab{a}})},\ \Eprint
  {http://arxiv.org/abs/1504.02100} {arXiv:1504.02100} \BibitemShut {NoStop}%
\bibitem [{\citenamefont {{Wang}}(2014)}]{Wang2014}%
  \BibitemOpen
  \bibfield  {author} {\bibinfo {author} {\bibfnamefont {Y.}~\bibnamefont
  {{Wang}}},\ }\href {\doibase 10.1093/mnras/stu1374} {\bibfield  {journal}
  {\bibinfo  {journal} {\mnras}\ }\textbf {\bibinfo {volume} {443}},\ \bibinfo
  {pages} {2950} (\bibinfo {year} {2014})},\ \Eprint
  {http://arxiv.org/abs/1404.5589} {arXiv:1404.5589} \BibitemShut {NoStop}%
\bibitem [{\citenamefont {{Shoji}}\ and\ \citenamefont
  {{Lee}}(2012)}]{Shoji2012}%
  \BibitemOpen
  \bibfield  {author} {\bibinfo {author} {\bibfnamefont {M.}~\bibnamefont
  {{Shoji}}}\ and\ \bibinfo {author} {\bibfnamefont {J.}~\bibnamefont
  {{Lee}}},\ }\href@noop {} {\bibfield  {journal} {\bibinfo  {journal} {ArXiv
  e-prints}\ } (\bibinfo {year} {2012})},\ \Eprint
  {http://arxiv.org/abs/1203.0869} {arXiv:1203.0869 [astro-ph.CO]} \BibitemShut
  {NoStop}%
\bibitem [{\citenamefont {{Paz}}\ \emph {et~al.}(2013)\citenamefont {{Paz}},
  \citenamefont {{Lares}}, \citenamefont {{Ceccarelli}}, \citenamefont
  {{Padilla}},\ and\ \citenamefont {{Lambas}}}]{Paz2013}%
  \BibitemOpen
  \bibfield  {author} {\bibinfo {author} {\bibfnamefont {D.}~\bibnamefont
  {{Paz}}}, \bibinfo {author} {\bibfnamefont {M.}~\bibnamefont {{Lares}}},
  \bibinfo {author} {\bibfnamefont {L.}~\bibnamefont {{Ceccarelli}}}, \bibinfo
  {author} {\bibfnamefont {N.}~\bibnamefont {{Padilla}}}, \ and\ \bibinfo
  {author} {\bibfnamefont {D.~G.}\ \bibnamefont {{Lambas}}},\ }\href {\doibase
  10.1093/mnras/stt1836} {\bibfield  {journal} {\bibinfo  {journal} {\mnras}\
  }\textbf {\bibinfo {volume} {436}},\ \bibinfo {pages} {3480} (\bibinfo {year}
  {2013})},\ \Eprint {http://arxiv.org/abs/1306.5799} {arXiv:1306.5799}
  \BibitemShut {NoStop}%
\bibitem [{\citenamefont {Hamaus}\ \emph {et~al.}(2016)\citenamefont {Hamaus},
  \citenamefont {Pisani}, \citenamefont {Sutter}, \citenamefont {Lavaux},
  \citenamefont {Escoffier}, \citenamefont {Wandelt},\ and\ \citenamefont
  {Weller}}]{Hamaus:2016wka}%
  \BibitemOpen
  \bibfield  {author} {\bibinfo {author} {\bibfnamefont {N.}~\bibnamefont
  {Hamaus}}, \bibinfo {author} {\bibfnamefont {A.}~\bibnamefont {Pisani}},
  \bibinfo {author} {\bibfnamefont {P.~M.}\ \bibnamefont {Sutter}}, \bibinfo
  {author} {\bibfnamefont {G.}~\bibnamefont {Lavaux}}, \bibinfo {author}
  {\bibfnamefont {S.}~\bibnamefont {Escoffier}}, \bibinfo {author}
  {\bibfnamefont {B.~D.}\ \bibnamefont {Wandelt}}, \ and\ \bibinfo {author}
  {\bibfnamefont {J.}~\bibnamefont {Weller}},\ }\href {\doibase
  10.1103/PhysRevLett.117.091302} {\bibfield  {journal} {\bibinfo  {journal}
  {Phys. Rev. Lett.}\ }\textbf {\bibinfo {volume} {117}},\ \bibinfo {pages}
  {091302} (\bibinfo {year} {2016})},\ \Eprint
  {http://arxiv.org/abs/1602.01784} {arXiv:1602.01784 [astro-ph.CO]}
  \BibitemShut {NoStop}%
%%CITATION = ARXIV:1602.01784;%%
\bibitem [{\citenamefont {Cai}\ \emph {et~al.}(2016)\citenamefont {Cai},
  \citenamefont {Taylor}, \citenamefont {Peacock},\ and\ \citenamefont
  {Padilla}}]{Cai:2016jek}%
  \BibitemOpen
  \bibfield  {author} {\bibinfo {author} {\bibfnamefont {Y.-C.}\ \bibnamefont
  {Cai}}, \bibinfo {author} {\bibfnamefont {A.}~\bibnamefont {Taylor}},
  \bibinfo {author} {\bibfnamefont {J.~A.}\ \bibnamefont {Peacock}}, \ and\
  \bibinfo {author} {\bibfnamefont {N.}~\bibnamefont {Padilla}},\ }\href
  {\doibase 10.1093/mnras/stw1809} {\  (\bibinfo {year} {2016}),\
  10.1093/mnras/stw1809},\ \Eprint {http://arxiv.org/abs/1603.05184}
  {arXiv:1603.05184 [astro-ph.CO]} \BibitemShut {NoStop}%
%%CITATION = ARXIV:1603.05184;%%
\bibitem [{\citenamefont {Achitouv}\ and\ \citenamefont
  {Blake}(2016)}]{Achitouv:2016mbn}%
  \BibitemOpen
  \bibfield  {author} {\bibinfo {author} {\bibfnamefont {I.}~\bibnamefont
  {Achitouv}}\ and\ \bibinfo {author} {\bibfnamefont {C.}~\bibnamefont
  {Blake}},\ }\href@noop {} {\  (\bibinfo {year} {2016})},\ \Eprint
  {http://arxiv.org/abs/1606.03092} {arXiv:1606.03092 [astro-ph.CO]}
  \BibitemShut {NoStop}%
%%CITATION = ARXIV:1606.03092;%%
\bibitem [{\citenamefont {{Seljak}}(2012)}]{Seljak2012}%
  \BibitemOpen
  \bibfield  {author} {\bibinfo {author} {\bibfnamefont {U.}~\bibnamefont
  {{Seljak}}},\ }\href {\doibase 10.1088/1475-7516/2012/03/004} {\bibfield
  {journal} {\bibinfo  {journal} {\jcap}\ }\textbf {\bibinfo {volume} {3}},\
  \bibinfo {eid} {004} (\bibinfo {year} {2012})},\ \Eprint
  {http://arxiv.org/abs/1201.0594} {arXiv:1201.0594} \BibitemShut {NoStop}%
\bibitem [{\citenamefont {{McDonald}}\ \emph {et~al.}(2000)\citenamefont
  {{McDonald}}, \citenamefont {{Miralda-Escud{\'e}}}, \citenamefont {{Rauch}},
  \citenamefont {{Sargent}}, \citenamefont {{Barlow}}, \citenamefont {{Cen}},\
  and\ \citenamefont {{Ostriker}}}]{McDonald2000}%
  \BibitemOpen
  \bibfield  {author} {\bibinfo {author} {\bibfnamefont {P.}~\bibnamefont
  {{McDonald}}}, \bibinfo {author} {\bibfnamefont {J.}~\bibnamefont
  {{Miralda-Escud{\'e}}}}, \bibinfo {author} {\bibfnamefont {M.}~\bibnamefont
  {{Rauch}}}, \bibinfo {author} {\bibfnamefont {W.~L.~W.}\ \bibnamefont
  {{Sargent}}}, \bibinfo {author} {\bibfnamefont {T.~A.}\ \bibnamefont
  {{Barlow}}}, \bibinfo {author} {\bibfnamefont {R.}~\bibnamefont {{Cen}}}, \
  and\ \bibinfo {author} {\bibfnamefont {J.~P.}\ \bibnamefont {{Ostriker}}},\
  }\href {\doibase 10.1086/317079} {\bibfield  {journal} {\bibinfo  {journal}
  {\apj}\ }\textbf {\bibinfo {volume} {543}},\ \bibinfo {pages} {1} (\bibinfo
  {year} {2000})},\ \Eprint {http://arxiv.org/abs/astro-ph/9911196}
  {astro-ph/9911196} \BibitemShut {NoStop}%
\bibitem [{\citenamefont {{McDonald}}(2003)}]{McDonald2003}%
  \BibitemOpen
  \bibfield  {author} {\bibinfo {author} {\bibfnamefont {P.}~\bibnamefont
  {{McDonald}}},\ }\href {\doibase 10.1086/345945} {\bibfield  {journal}
  {\bibinfo  {journal} {\apj}\ }\textbf {\bibinfo {volume} {585}},\ \bibinfo
  {pages} {34} (\bibinfo {year} {2003})},\ \Eprint
  {http://arxiv.org/abs/astro-ph/0108064} {astro-ph/0108064} \BibitemShut
  {NoStop}%
\bibitem [{\citenamefont {{Wang}}\ \emph {et~al.}(2015)\citenamefont {{Wang}},
  \citenamefont {{Font-Ribera}},\ and\ \citenamefont {{Seljak}}}]{Wang2015}%
  \BibitemOpen
  \bibfield  {author} {\bibinfo {author} {\bibfnamefont {X.}~\bibnamefont
  {{Wang}}}, \bibinfo {author} {\bibfnamefont {A.}~\bibnamefont
  {{Font-Ribera}}}, \ and\ \bibinfo {author} {\bibfnamefont {U.}~\bibnamefont
  {{Seljak}}},\ }\href {\doibase 10.1088/1475-7516/2015/04/009} {\bibfield
  {journal} {\bibinfo  {journal} {\jcap}\ }\textbf {\bibinfo {volume} {4}},\
  \bibinfo {eid} {009} (\bibinfo {year} {2015})},\ \Eprint
  {http://arxiv.org/abs/1412.4727} {arXiv:1412.4727} \BibitemShut {NoStop}%
\bibitem [{\citenamefont {{Kaiser}}(1984)}]{Kaiser1984}%
  \BibitemOpen
  \bibfield  {author} {\bibinfo {author} {\bibfnamefont {N.}~\bibnamefont
  {{Kaiser}}},\ }\href {\doibase 10.1086/184341} {\bibfield  {journal}
  {\bibinfo  {journal} {\apjl}\ }\textbf {\bibinfo {volume} {284}},\ \bibinfo
  {pages} {L9} (\bibinfo {year} {1984})}\BibitemShut {NoStop}%
\bibitem [{\citenamefont {{Coles}}(1993)}]{Coles1993}%
  \BibitemOpen
  \bibfield  {author} {\bibinfo {author} {\bibfnamefont {P.}~\bibnamefont
  {{Coles}}},\ }\href {\doibase 10.1093/mnras/262.4.1065} {\bibfield  {journal}
  {\bibinfo  {journal} {\mnras}\ }\textbf {\bibinfo {volume} {262}},\ \bibinfo
  {pages} {1065} (\bibinfo {year} {1993})}\BibitemShut {NoStop}%
\bibitem [{\citenamefont {{Fry}}\ and\ \citenamefont
  {{Gaztanaga}}(1993)}]{Fry1993}%
  \BibitemOpen
  \bibfield  {author} {\bibinfo {author} {\bibfnamefont {J.~N.}\ \bibnamefont
  {{Fry}}}\ and\ \bibinfo {author} {\bibfnamefont {E.}~\bibnamefont
  {{Gaztanaga}}},\ }\href {\doibase 10.1086/173015} {\bibfield  {journal}
  {\bibinfo  {journal} {\apj}\ }\textbf {\bibinfo {volume} {413}},\ \bibinfo
  {pages} {447} (\bibinfo {year} {1993})},\ \Eprint
  {http://arxiv.org/abs/astro-ph/9302009} {astro-ph/9302009} \BibitemShut
  {NoStop}%
\bibitem [{\citenamefont {{Bond}}\ and\ \citenamefont
  {{Myers}}(1996)}]{Bond1996}%
  \BibitemOpen
  \bibfield  {author} {\bibinfo {author} {\bibfnamefont {J.~R.}\ \bibnamefont
  {{Bond}}}\ and\ \bibinfo {author} {\bibfnamefont {S.~T.}\ \bibnamefont
  {{Myers}}},\ }\href {\doibase 10.1086/192269} {\bibfield  {journal} {\bibinfo
   {journal} {\apjs}\ }\textbf {\bibinfo {volume} {103}},\ \bibinfo {pages}
  {63} (\bibinfo {year} {1996})}\BibitemShut {NoStop}%
\bibitem [{\citenamefont {{Dekel}}\ and\ \citenamefont
  {{Lahav}}(1999)}]{Dekel1999}%
  \BibitemOpen
  \bibfield  {author} {\bibinfo {author} {\bibfnamefont {A.}~\bibnamefont
  {{Dekel}}}\ and\ \bibinfo {author} {\bibfnamefont {O.}~\bibnamefont
  {{Lahav}}},\ }\href {\doibase 10.1086/307428} {\bibfield  {journal} {\bibinfo
   {journal} {\apj}\ }\textbf {\bibinfo {volume} {520}},\ \bibinfo {pages} {24}
  (\bibinfo {year} {1999})},\ \Eprint {http://arxiv.org/abs/astro-ph/9806193}
  {astro-ph/9806193} \BibitemShut {NoStop}%
\bibitem [{\citenamefont {{Seljak}}(2000)}]{Seljak2000}%
  \BibitemOpen
  \bibfield  {author} {\bibinfo {author} {\bibfnamefont {U.}~\bibnamefont
  {{Seljak}}},\ }\href {\doibase 10.1046/j.1365-8711.2000.03715.x} {\bibfield
  {journal} {\bibinfo  {journal} {\mnras}\ }\textbf {\bibinfo {volume} {318}},\
  \bibinfo {pages} {203} (\bibinfo {year} {2000})},\ \Eprint
  {http://arxiv.org/abs/astro-ph/0001493} {astro-ph/0001493} \BibitemShut
  {NoStop}%
\bibitem [{\citenamefont {{Berlind}}\ \emph {et~al.}(2003)\citenamefont
  {{Berlind}}, \citenamefont {{Weinberg}}, \citenamefont {{Benson}},
  \citenamefont {{Baugh}}, \citenamefont {{Cole}}, \citenamefont {{Dav{\'e}}},
  \citenamefont {{Frenk}}, \citenamefont {{Jenkins}}, \citenamefont {{Katz}},\
  and\ \citenamefont {{Lacey}}}]{Berlind2003}%
  \BibitemOpen
  \bibfield  {author} {\bibinfo {author} {\bibfnamefont {A.~A.}\ \bibnamefont
  {{Berlind}}}, \bibinfo {author} {\bibfnamefont {D.~H.}\ \bibnamefont
  {{Weinberg}}}, \bibinfo {author} {\bibfnamefont {A.~J.}\ \bibnamefont
  {{Benson}}}, \bibinfo {author} {\bibfnamefont {C.~M.}\ \bibnamefont
  {{Baugh}}}, \bibinfo {author} {\bibfnamefont {S.}~\bibnamefont {{Cole}}},
  \bibinfo {author} {\bibfnamefont {R.}~\bibnamefont {{Dav{\'e}}}}, \bibinfo
  {author} {\bibfnamefont {C.~S.}\ \bibnamefont {{Frenk}}}, \bibinfo {author}
  {\bibfnamefont {A.}~\bibnamefont {{Jenkins}}}, \bibinfo {author}
  {\bibfnamefont {N.}~\bibnamefont {{Katz}}}, \ and\ \bibinfo {author}
  {\bibfnamefont {C.~G.}\ \bibnamefont {{Lacey}}},\ }\href {\doibase
  10.1086/376517} {\bibfield  {journal} {\bibinfo  {journal} {\apj}\ }\textbf
  {\bibinfo {volume} {593}},\ \bibinfo {pages} {1} (\bibinfo {year} {2003})},\
  \Eprint {http://arxiv.org/abs/astro-ph/0212357} {astro-ph/0212357}
  \BibitemShut {NoStop}%
\bibitem [{\citenamefont {{McDonald}}\ and\ \citenamefont
  {{Roy}}(2009)}]{McDonald2009}%
  \BibitemOpen
  \bibfield  {author} {\bibinfo {author} {\bibfnamefont {P.}~\bibnamefont
  {{McDonald}}}\ and\ \bibinfo {author} {\bibfnamefont {A.}~\bibnamefont
  {{Roy}}},\ }\href {\doibase 10.1088/1475-7516/2009/08/020} {\bibfield
  {journal} {\bibinfo  {journal} {\jcap}\ }\textbf {\bibinfo {volume} {8}},\
  \bibinfo {eid} {020} (\bibinfo {year} {2009})},\ \Eprint
  {http://arxiv.org/abs/0902.0991} {arXiv:0902.0991 [astro-ph.CO]} \BibitemShut
  {NoStop}%
\bibitem [{\citenamefont {{Desjacques}}\ \emph {et~al.}(2010)\citenamefont
  {{Desjacques}}, \citenamefont {{Crocce}}, \citenamefont {{Scoccimarro}},\
  and\ \citenamefont {{Sheth}}}]{Desjacques2010}%
  \BibitemOpen
  \bibfield  {author} {\bibinfo {author} {\bibfnamefont {V.}~\bibnamefont
  {{Desjacques}}}, \bibinfo {author} {\bibfnamefont {M.}~\bibnamefont
  {{Crocce}}}, \bibinfo {author} {\bibfnamefont {R.}~\bibnamefont
  {{Scoccimarro}}}, \ and\ \bibinfo {author} {\bibfnamefont {R.~K.}\
  \bibnamefont {{Sheth}}},\ }\href {\doibase 10.1103/PhysRevD.82.103529}
  {\bibfield  {journal} {\bibinfo  {journal} {\prd}\ }\textbf {\bibinfo
  {volume} {82}},\ \bibinfo {eid} {103529} (\bibinfo {year} {2010})},\ \Eprint
  {http://arxiv.org/abs/1009.3449} {arXiv:1009.3449 [astro-ph.CO]} \BibitemShut
  {NoStop}%
\bibitem [{\citenamefont {{Matsubara}}(2011)}]{Matsubara2011}%
  \BibitemOpen
  \bibfield  {author} {\bibinfo {author} {\bibfnamefont {T.}~\bibnamefont
  {{Matsubara}}},\ }\href {\doibase 10.1103/PhysRevD.83.083518} {\bibfield
  {journal} {\bibinfo  {journal} {\prd}\ }\textbf {\bibinfo {volume} {83}},\
  \bibinfo {eid} {083518} (\bibinfo {year} {2011})},\ \Eprint
  {http://arxiv.org/abs/1102.4619} {arXiv:1102.4619 [astro-ph.CO]} \BibitemShut
  {NoStop}%
\bibitem [{\citenamefont {{Schmidt}}(2013)}]{Schmidt2013b}%
  \BibitemOpen
  \bibfield  {author} {\bibinfo {author} {\bibfnamefont {F.}~\bibnamefont
  {{Schmidt}}},\ }\href {\doibase 10.1103/PhysRevD.87.123518} {\bibfield
  {journal} {\bibinfo  {journal} {\prd}\ }\textbf {\bibinfo {volume} {87}},\
  \bibinfo {eid} {123518} (\bibinfo {year} {2013})},\ \Eprint
  {http://arxiv.org/abs/1304.1817} {arXiv:1304.1817 [astro-ph.CO]} \BibitemShut
  {NoStop}%
\bibitem [{\citenamefont {{Kitaura}}\ \emph {et~al.}(2014)\citenamefont
  {{Kitaura}}, \citenamefont {{Yepes}},\ and\ \citenamefont
  {{Prada}}}]{Kitaura2014}%
  \BibitemOpen
  \bibfield  {author} {\bibinfo {author} {\bibfnamefont {F.-S.}\ \bibnamefont
  {{Kitaura}}}, \bibinfo {author} {\bibfnamefont {G.}~\bibnamefont {{Yepes}}},
  \ and\ \bibinfo {author} {\bibfnamefont {F.}~\bibnamefont {{Prada}}},\ }\href
  {\doibase 10.1093/mnrasl/slt172} {\bibfield  {journal} {\bibinfo  {journal}
  {\mnras}\ }\textbf {\bibinfo {volume} {439}},\ \bibinfo {pages} {L21}
  (\bibinfo {year} {2014})},\ \Eprint {http://arxiv.org/abs/1307.3285}
  {arXiv:1307.3285} \BibitemShut {NoStop}%
\bibitem [{\citenamefont {{Saito}}\ \emph {et~al.}(2014)\citenamefont
  {{Saito}}, \citenamefont {{Baldauf}}, \citenamefont {{Vlah}}, \citenamefont
  {{Seljak}}, \citenamefont {{Okumura}},\ and\ \citenamefont
  {{McDonald}}}]{Saito2014}%
  \BibitemOpen
  \bibfield  {author} {\bibinfo {author} {\bibfnamefont {S.}~\bibnamefont
  {{Saito}}}, \bibinfo {author} {\bibfnamefont {T.}~\bibnamefont {{Baldauf}}},
  \bibinfo {author} {\bibfnamefont {Z.}~\bibnamefont {{Vlah}}}, \bibinfo
  {author} {\bibfnamefont {U.}~\bibnamefont {{Seljak}}}, \bibinfo {author}
  {\bibfnamefont {T.}~\bibnamefont {{Okumura}}}, \ and\ \bibinfo {author}
  {\bibfnamefont {P.}~\bibnamefont {{McDonald}}},\ }\href {\doibase
  10.1103/PhysRevD.90.123522} {\bibfield  {journal} {\bibinfo  {journal}
  {\prd}\ }\textbf {\bibinfo {volume} {90}},\ \bibinfo {eid} {123522} (\bibinfo
  {year} {2014})},\ \Eprint {http://arxiv.org/abs/1405.1447} {arXiv:1405.1447}
  \BibitemShut {NoStop}%
\bibitem [{\citenamefont {{Baldauf}}\ \emph {et~al.}(2015)\citenamefont
  {{Baldauf}}, \citenamefont {{Mercolli}},\ and\ \citenamefont
  {{Zaldarriaga}}}]{Baldauf2015}%
  \BibitemOpen
  \bibfield  {author} {\bibinfo {author} {\bibfnamefont {T.}~\bibnamefont
  {{Baldauf}}}, \bibinfo {author} {\bibfnamefont {L.}~\bibnamefont
  {{Mercolli}}}, \ and\ \bibinfo {author} {\bibfnamefont {M.}~\bibnamefont
  {{Zaldarriaga}}},\ }\href {\doibase 10.1103/PhysRevD.92.123007} {\bibfield
  {journal} {\bibinfo  {journal} {\prd}\ }\textbf {\bibinfo {volume} {92}},\
  \bibinfo {eid} {123007} (\bibinfo {year} {2015})},\ \Eprint
  {http://arxiv.org/abs/1507.02256} {arXiv:1507.02256} \BibitemShut {NoStop}%
\bibitem [{\citenamefont {Zhao}\ \emph {et~al.}(2016)\citenamefont {Zhao},
  \citenamefont {Tao}, \citenamefont {Liang}, \citenamefont {Kitaura},\ and\
  \citenamefont {Chuang}}]{Zhao:2015ecx}%
  \BibitemOpen
  \bibfield  {author} {\bibinfo {author} {\bibfnamefont {C.}~\bibnamefont
  {Zhao}}, \bibinfo {author} {\bibfnamefont {C.}~\bibnamefont {Tao}}, \bibinfo
  {author} {\bibfnamefont {Y.}~\bibnamefont {Liang}}, \bibinfo {author}
  {\bibfnamefont {F.-S.}\ \bibnamefont {Kitaura}}, \ and\ \bibinfo {author}
  {\bibfnamefont {C.-H.}\ \bibnamefont {Chuang}},\ }\href {\doibase
  10.1093/mnras/stw660} {\bibfield  {journal} {\bibinfo  {journal} {Mon. Not.
  Roy. Astron. Soc.}\ }\textbf {\bibinfo {volume} {459}},\ \bibinfo {pages}
  {2670} (\bibinfo {year} {2016})},\ \Eprint {http://arxiv.org/abs/1511.04299}
  {arXiv:1511.04299 [astro-ph.CO]} \BibitemShut {NoStop}%
%%CITATION = ARXIV:1511.04299;%%
\bibitem [{\citenamefont {{McDonald}}(2006)}]{McDonald2006}%
  \BibitemOpen
  \bibfield  {author} {\bibinfo {author} {\bibfnamefont {P.}~\bibnamefont
  {{McDonald}}},\ }\href {\doibase 10.1103/PhysRevD.74.103512} {\bibfield
  {journal} {\bibinfo  {journal} {\prd}\ }\textbf {\bibinfo {volume} {74}},\
  \bibinfo {eid} {103512} (\bibinfo {year} {2006})},\ \Eprint
  {http://arxiv.org/abs/astro-ph/0609413} {astro-ph/0609413} \BibitemShut
  {NoStop}%
\bibitem [{\citenamefont {{Alam}}\ \emph
  {et~al.}(2015{\natexlab{b}})\citenamefont {{Alam}}, \citenamefont
  {{Albareti}}, \citenamefont {{Allende Prieto}}, \citenamefont {{Anders}},
  \citenamefont {{Anderson}}, \citenamefont {{Anderton}}, \citenamefont
  {{Andrews}}, \citenamefont {{Armengaud}}, \citenamefont {{Aubourg}},
  \citenamefont {{Bailey}},\ and\ \citenamefont {et~al.}}]{Alam15}%
  \BibitemOpen
  \bibfield  {author} {\bibinfo {author} {\bibfnamefont {S.}~\bibnamefont
  {{Alam}}}, \bibinfo {author} {\bibfnamefont {F.~D.}\ \bibnamefont
  {{Albareti}}}, \bibinfo {author} {\bibfnamefont {C.}~\bibnamefont {{Allende
  Prieto}}}, \bibinfo {author} {\bibfnamefont {F.}~\bibnamefont {{Anders}}},
  \bibinfo {author} {\bibfnamefont {S.~F.}\ \bibnamefont {{Anderson}}},
  \bibinfo {author} {\bibfnamefont {T.}~\bibnamefont {{Anderton}}}, \bibinfo
  {author} {\bibfnamefont {B.~H.}\ \bibnamefont {{Andrews}}}, \bibinfo {author}
  {\bibfnamefont {E.}~\bibnamefont {{Armengaud}}}, \bibinfo {author}
  {\bibfnamefont {{\'E}.}~\bibnamefont {{Aubourg}}}, \bibinfo {author}
  {\bibfnamefont {S.}~\bibnamefont {{Bailey}}}, \ and\ \bibinfo {author}
  {\bibnamefont {et~al.}},\ }\href {\doibase 10.1088/0067-0049/219/1/12}
  {\bibfield  {journal} {\bibinfo  {journal} {\apjs}\ }\textbf {\bibinfo
  {volume} {219}},\ \bibinfo {eid} {12} (\bibinfo {year}
  {2015}{\natexlab{b}})},\ \Eprint {http://arxiv.org/abs/1501.00963}
  {arXiv:1501.00963 [astro-ph.IM]} \BibitemShut {NoStop}%
\bibitem [{\citenamefont {{Eisenstein}}\ \emph {et~al.}(2011)\citenamefont
  {{Eisenstein}}, \citenamefont {{Weinberg}}, \citenamefont {{Agol}},
  \citenamefont {{Aihara}}, \citenamefont {{Allende Prieto}}, \citenamefont
  {{Anderson}}, \citenamefont {{Arns}}, \citenamefont {{Aubourg}},
  \citenamefont {{Bailey}}, \citenamefont {{Balbinot}},\ and\ \citenamefont
  {et~al.}}]{Einstein2011}%
  \BibitemOpen
  \bibfield  {author} {\bibinfo {author} {\bibfnamefont {D.~J.}\ \bibnamefont
  {{Eisenstein}}}, \bibinfo {author} {\bibfnamefont {D.~H.}\ \bibnamefont
  {{Weinberg}}}, \bibinfo {author} {\bibfnamefont {E.}~\bibnamefont {{Agol}}},
  \bibinfo {author} {\bibfnamefont {H.}~\bibnamefont {{Aihara}}}, \bibinfo
  {author} {\bibfnamefont {C.}~\bibnamefont {{Allende Prieto}}}, \bibinfo
  {author} {\bibfnamefont {S.~F.}\ \bibnamefont {{Anderson}}}, \bibinfo
  {author} {\bibfnamefont {J.~A.}\ \bibnamefont {{Arns}}}, \bibinfo {author}
  {\bibfnamefont {{\'E}.}~\bibnamefont {{Aubourg}}}, \bibinfo {author}
  {\bibfnamefont {S.}~\bibnamefont {{Bailey}}}, \bibinfo {author}
  {\bibfnamefont {E.}~\bibnamefont {{Balbinot}}}, \ and\ \bibinfo {author}
  {\bibnamefont {et~al.}},\ }\href {\doibase 10.1088/0004-6256/142/3/72}
  {\bibfield  {journal} {\bibinfo  {journal} {\aj}\ }\textbf {\bibinfo {volume}
  {142}},\ \bibinfo {eid} {72} (\bibinfo {year} {2011})},\ \Eprint
  {http://arxiv.org/abs/1101.1529} {arXiv:1101.1529 [astro-ph.IM]} \BibitemShut
  {NoStop}%
\bibitem [{\citenamefont {{Gunn}}\ \emph {et~al.}(2006)\citenamefont {{Gunn}},
  \citenamefont {{Siegmund}}, \citenamefont {{Mannery}}, \citenamefont
  {{Owen}}, \citenamefont {{Hull}}, \citenamefont {{Leger}}, \citenamefont
  {{Carey}}, \citenamefont {{Knapp}},\ and\ \citenamefont {{et
  al.,}}}]{Gunn2006}%
  \BibitemOpen
  \bibfield  {author} {\bibinfo {author} {\bibfnamefont {J.~E.}\ \bibnamefont
  {{Gunn}}}, \bibinfo {author} {\bibfnamefont {W.~A.}\ \bibnamefont
  {{Siegmund}}}, \bibinfo {author} {\bibfnamefont {E.~J.}\ \bibnamefont
  {{Mannery}}}, \bibinfo {author} {\bibfnamefont {R.~E.}\ \bibnamefont
  {{Owen}}}, \bibinfo {author} {\bibfnamefont {C.~L.}\ \bibnamefont {{Hull}}},
  \bibinfo {author} {\bibfnamefont {R.~F.}\ \bibnamefont {{Leger}}}, \bibinfo
  {author} {\bibfnamefont {L.~N.}\ \bibnamefont {{Carey}}}, \bibinfo {author}
  {\bibfnamefont {G.~R.}\ \bibnamefont {{Knapp}}}, \ and\ \bibinfo {author}
  {\bibnamefont {{et al.,}}},\ }\href {\doibase 10.1086/500975} {\bibfield
  {journal} {\bibinfo  {journal} {\aj}\ }\textbf {\bibinfo {volume} {131}},\
  \bibinfo {pages} {2332} (\bibinfo {year} {2006})},\ \Eprint
  {http://arxiv.org/abs/arXiv:astro-ph/0602326} {arXiv:astro-ph/0602326}
  \BibitemShut {NoStop}%
\bibitem [{\citenamefont {{Smee}}\ \emph {et~al.}(2013)\citenamefont {{Smee}},
  \citenamefont {{Gunn}}, \citenamefont {{Uomoto}}, \citenamefont {{Roe}},
  \citenamefont {{Schlegel}}, \citenamefont {{Rockosi}}, \citenamefont
  {{Carr}}, \citenamefont {{Leger}},\ and\ \citenamefont {et~al.}}]{Smee2013}%
  \BibitemOpen
  \bibfield  {author} {\bibinfo {author} {\bibfnamefont {S.~A.}\ \bibnamefont
  {{Smee}}}, \bibinfo {author} {\bibfnamefont {J.~E.}\ \bibnamefont {{Gunn}}},
  \bibinfo {author} {\bibfnamefont {A.}~\bibnamefont {{Uomoto}}}, \bibinfo
  {author} {\bibfnamefont {N.}~\bibnamefont {{Roe}}}, \bibinfo {author}
  {\bibfnamefont {D.}~\bibnamefont {{Schlegel}}}, \bibinfo {author}
  {\bibfnamefont {C.~M.}\ \bibnamefont {{Rockosi}}}, \bibinfo {author}
  {\bibfnamefont {M.~A.}\ \bibnamefont {{Carr}}}, \bibinfo {author}
  {\bibfnamefont {F.}~\bibnamefont {{Leger}}}, \ and\ \bibinfo {author}
  {\bibnamefont {et~al.}},\ }\href {\doibase 10.1088/0004-6256/146/2/32}
  {\bibfield  {journal} {\bibinfo  {journal} {\aj}\ }\textbf {\bibinfo {volume}
  {146}},\ \bibinfo {eid} {32} (\bibinfo {year} {2013})},\ \Eprint
  {http://arxiv.org/abs/1208.2233} {arXiv:1208.2233 [astro-ph.IM]} \BibitemShut
  {NoStop}%
\bibitem [{\citenamefont {{Bolton}}\ \emph {et~al.}(2012)\citenamefont
  {{Bolton}}, \citenamefont {{Schlegel}}, \citenamefont {{Aubourg}},
  \citenamefont {{Bailey}}, \citenamefont {{Bhardwaj}}, \citenamefont
  {{Brownstein}}, \citenamefont {{Burles}}, \citenamefont {{Chen}},\ and\
  \citenamefont {{et al.}}}]{Blanton2012}%
  \BibitemOpen
  \bibfield  {author} {\bibinfo {author} {\bibfnamefont {A.~S.}\ \bibnamefont
  {{Bolton}}}, \bibinfo {author} {\bibfnamefont {D.~J.}\ \bibnamefont
  {{Schlegel}}}, \bibinfo {author} {\bibfnamefont {{\'E}.}~\bibnamefont
  {{Aubourg}}}, \bibinfo {author} {\bibfnamefont {S.}~\bibnamefont {{Bailey}}},
  \bibinfo {author} {\bibfnamefont {V.}~\bibnamefont {{Bhardwaj}}}, \bibinfo
  {author} {\bibfnamefont {J.~R.}\ \bibnamefont {{Brownstein}}}, \bibinfo
  {author} {\bibfnamefont {S.}~\bibnamefont {{Burles}}}, \bibinfo {author}
  {\bibfnamefont {Y.-M.}\ \bibnamefont {{Chen}}}, \ and\ \bibinfo {author}
  {\bibnamefont {{et al.}}},\ }\href {\doibase 10.1088/0004-6256/144/5/144}
  {\bibfield  {journal} {\bibinfo  {journal} {\aj}\ }\textbf {\bibinfo {volume}
  {144}},\ \bibinfo {eid} {144} (\bibinfo {year} {2012})},\ \Eprint
  {http://arxiv.org/abs/1207.7326} {arXiv:1207.7326} \BibitemShut {NoStop}%
\bibitem [{\citenamefont {{Reid}}\ and\ \citenamefont {{et
  al.}}(2016)}]{Reidetal2016}%
  \BibitemOpen
  \bibfield  {author} {\bibinfo {author} {\bibfnamefont {B.}~\bibnamefont
  {{Reid}}}\ and\ \bibinfo {author} {\bibnamefont {{et al.}}},\ }\href
  {\doibase 10.1093/mnras/stv2382} {\bibfield  {journal} {\bibinfo  {journal}
  {\mnras}\ }\textbf {\bibinfo {volume} {455}},\ \bibinfo {pages} {1553}
  (\bibinfo {year} {2016})},\ \Eprint {http://arxiv.org/abs/1509.06529}
  {arXiv:1509.06529} \BibitemShut {NoStop}%
\bibitem [{\citenamefont {{Anderson}}\ and\ \citenamefont {{et
  al.}}(2014)}]{Andersonetal2014}%
  \BibitemOpen
  \bibfield  {author} {\bibinfo {author} {\bibfnamefont {L.}~\bibnamefont
  {{Anderson}}}\ and\ \bibinfo {author} {\bibnamefont {{et al.}}},\ }\href
  {\doibase 10.1093/mnras/stu523} {\bibfield  {journal} {\bibinfo  {journal}
  {\mnras}\ }\textbf {\bibinfo {volume} {441}},\ \bibinfo {pages} {24}
  (\bibinfo {year} {2014})},\ \Eprint {http://arxiv.org/abs/1312.4877}
  {arXiv:1312.4877} \BibitemShut {NoStop}%
\bibitem [{\citenamefont {{Kitaura et al.}}\ \emph {et~al.}(2016)\citenamefont
  {{Kitaura et al.}}, \citenamefont {{Rodr{\'{\i}}guez-Torres}}, \citenamefont
  {{Chuang}}, \citenamefont {{Zhao}},\ and\ \citenamefont {{et
  al.}}}]{Kitauraetal2016a}%
  \BibitemOpen
  \bibfield  {author} {\bibinfo {author} {\bibfnamefont {F.-S.}\ \bibnamefont
  {{Kitaura et al.}}}, \bibinfo {author} {\bibfnamefont {S.}~\bibnamefont
  {{Rodr{\'{\i}}guez-Torres}}}, \bibinfo {author} {\bibfnamefont {C.-H.}\
  \bibnamefont {{Chuang}}}, \bibinfo {author} {\bibfnamefont {C.}~\bibnamefont
  {{Zhao}}}, \ and\ \bibinfo {author} {\bibnamefont {{et al.}}},\ }\href
  {\doibase 10.1093/mnras/stv2826} {\bibfield  {journal} {\bibinfo  {journal}
  {\mnras}\ }\textbf {\bibinfo {volume} {456}},\ \bibinfo {pages} {4156}
  (\bibinfo {year} {2016})},\ \Eprint {http://arxiv.org/abs/1509.06400}
  {arXiv:1509.06400} \BibitemShut {NoStop}%
\bibitem [{\citenamefont {Rodr{\'\i}guez-Torres}\ \emph
  {et~al.}(2015)\citenamefont {Rodr{\'\i}guez-Torres} \emph
  {et~al.}}]{Rodriguez-Torres:2015vqa}%
  \BibitemOpen
  \bibfield  {author} {\bibinfo {author} {\bibfnamefont {S.~A.}\ \bibnamefont
  {Rodr{\'\i}guez-Torres}} \emph {et~al.},\ }\href {\doibase
  10.1093/mnras/stw1014} {\  (\bibinfo {year} {2015}),\
  10.1093/mnras/stw1014},\ \Eprint {http://arxiv.org/abs/1509.06404}
  {arXiv:1509.06404 [astro-ph.CO]} \BibitemShut {NoStop}%
%%CITATION = ARXIV:1509.06404;%%
\end{thebibliography}%

%\begin{thebibliography}{99}
%\end{thebibliography}

%%%%%%%%%%%%%%%%%%%%%%%%%%%%%%%%%%%%%%%%%%%%%%%%%%

 \appendix

\section{Particular case: zero bias voids}

In this section we use data from the Data Release DR11 (see Ref.~\citep[][]{Alam15}) of the Baryon Oscillation Spectroscopic Survey  (BOSS, see Ref.~\citep[][]{Einstein2011}). The BOSS survey uses the SDSS 2.5 meter telescope at Apache Point Observatory (see Ref.~\citep[][]{Gunn2006}) and the spectra are obtained using the double-armed BOSS spectrograph (see Ref.~\citep[][]{Smee2013}). The data are then reduced using the algorithms described in \citep[][]{Blanton2012}.  The target selection of the CMASS and LOWZ samples, together with the algorithms used to create large scale structure catalogs (the \textsc{mksample} code), are presented in  Ref.~\citep[][]{Reidetal2016}.

We restrict this analysis to the CMASS sample of luminous red galaxies (LRGs), which is a complete sample, nearly constant in mass and volume limited between the redshifts $0.43\le z \le 0.7$ (see \citep{Reidetal2016,Andersonetal2014} for details of the targeting strategy).

Based on the mock galaxy catalogs for the CMASS sample (see Ref.~\citep[][]{Kitauraetal2016a,Rodriguez-Torres:2015vqa})  and on the void catalog obtained with the \textsc{dive} code (see Ref.~\citep[][]{Zhao:2015ecx}) we compute the quadrupoles for the void population selected with a radius cut of 16  $h^{-1}$ Mpc (see Fig.~\ref{fig:quad-cmass}). This is the population leading to the largest BAO signal-to-noise ratio without further considering optimal weights (see Ref.~\citep[][]{Liang2016}) used to measure the BAO from CMASS BOSS DR11 data (see Ref.~\citep[][]{Kitauraetal2016b}).
 We find a closely vanishing quadrupole at large scales. We see a similar behavior from DR11 \textsc{patchy} mock catalogs. 

In Fig.~\ref{fig:2dcf} we can see that the 2D correlation functions from the observed void  catalog is as compared to the galaxies. We find that the correlation function vanishes on large scales, as expected for zero bias tracers.
This particular case gives further support to the void bias model being tracer of the linear galaxy redshift space field.

\bfiw
\begin{tabular}{cc}
\includegraphics[width=10.3cm]{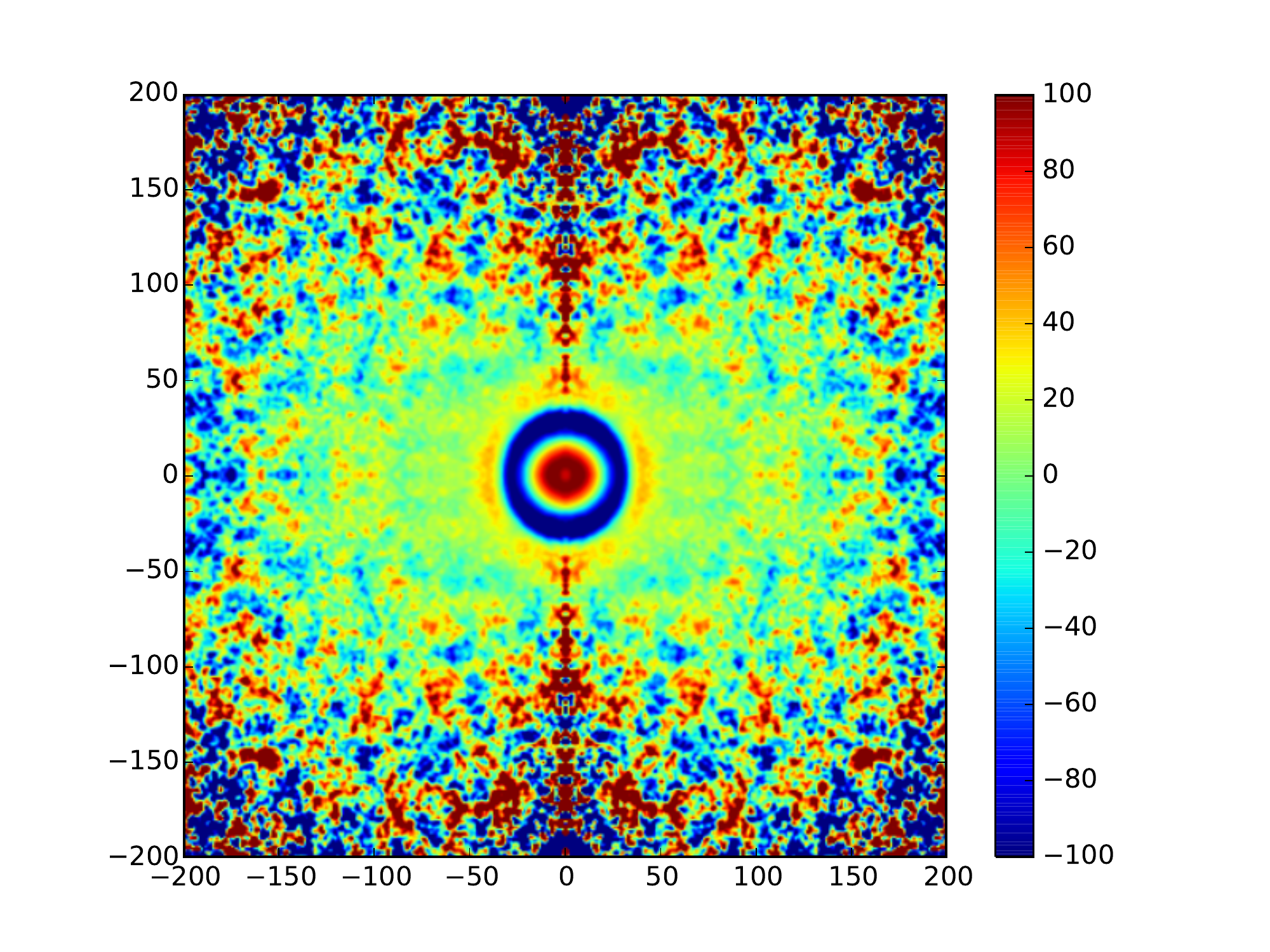}
\put(-170,210){\rotatebox[]{0}{$s^2\,\xi^{{\rm v}{\rm v}}(\sigma,\pi)$}}
\put(-280,108){\rotatebox[]{90}{$\pi$}}
\put(-165,0){\rotatebox[]{0}{$\sigma$}}
\hspace{-2.5cm}
\includegraphics[width=10.3cm]{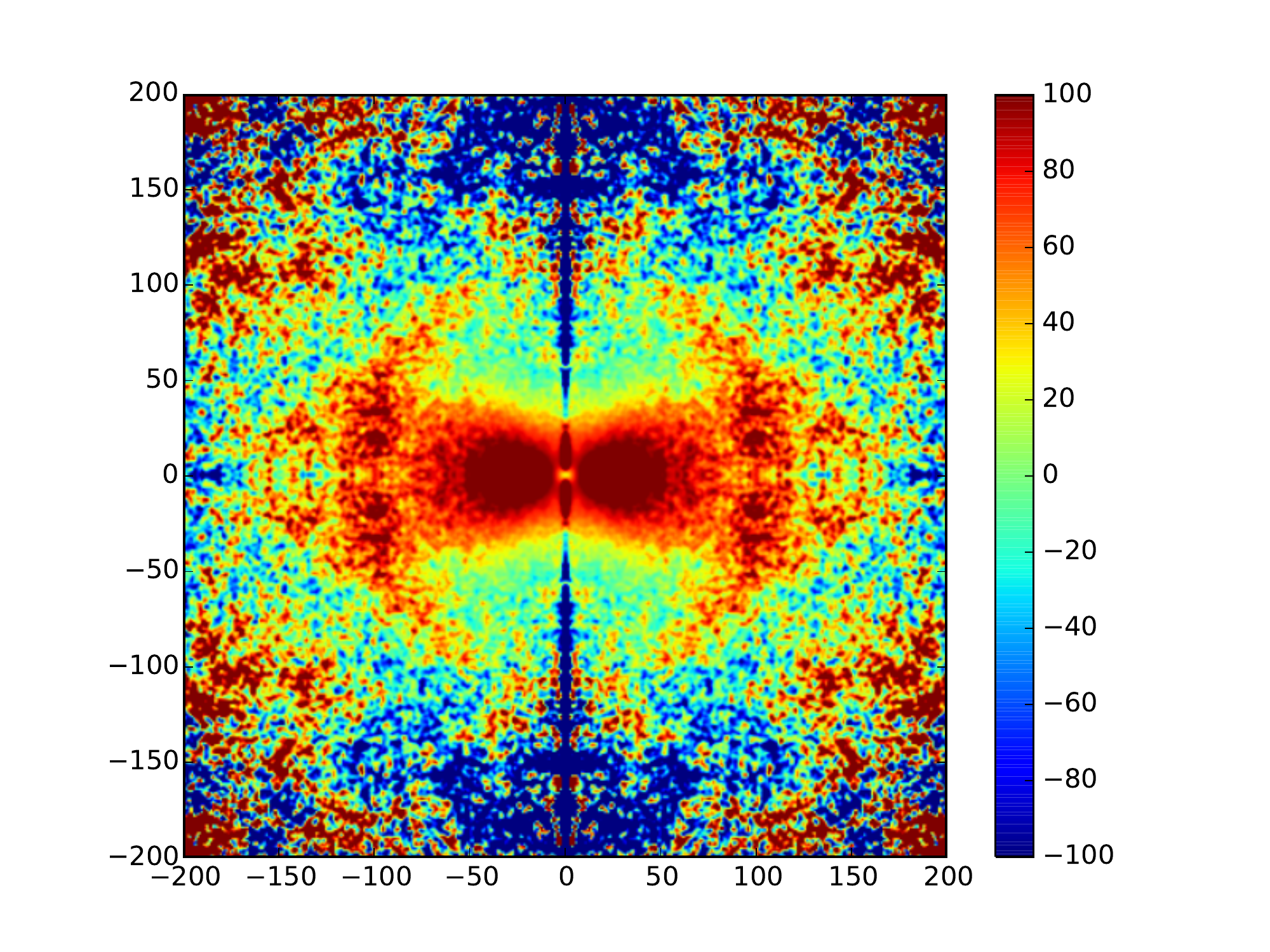}
\put(-170,210){\rotatebox[]{0}{$s^2\,\xi^{{\rm g}{\rm g}}(\sigma,\pi)$}}
\put(-280,108){\rotatebox[]{90}{$\pi$}}
\put(-165,0){\rotatebox[]{0}{$\sigma$}}
\end{tabular}
\caption{2D correlation function for CMASS DR11 left panel:   voids; right panel:  galaxies. }
\label{fig:2dcf}
\efiw

\bfi
\begin{center}
\hspace{-0.5cm}
\includegraphics[width=.5\textwidth]{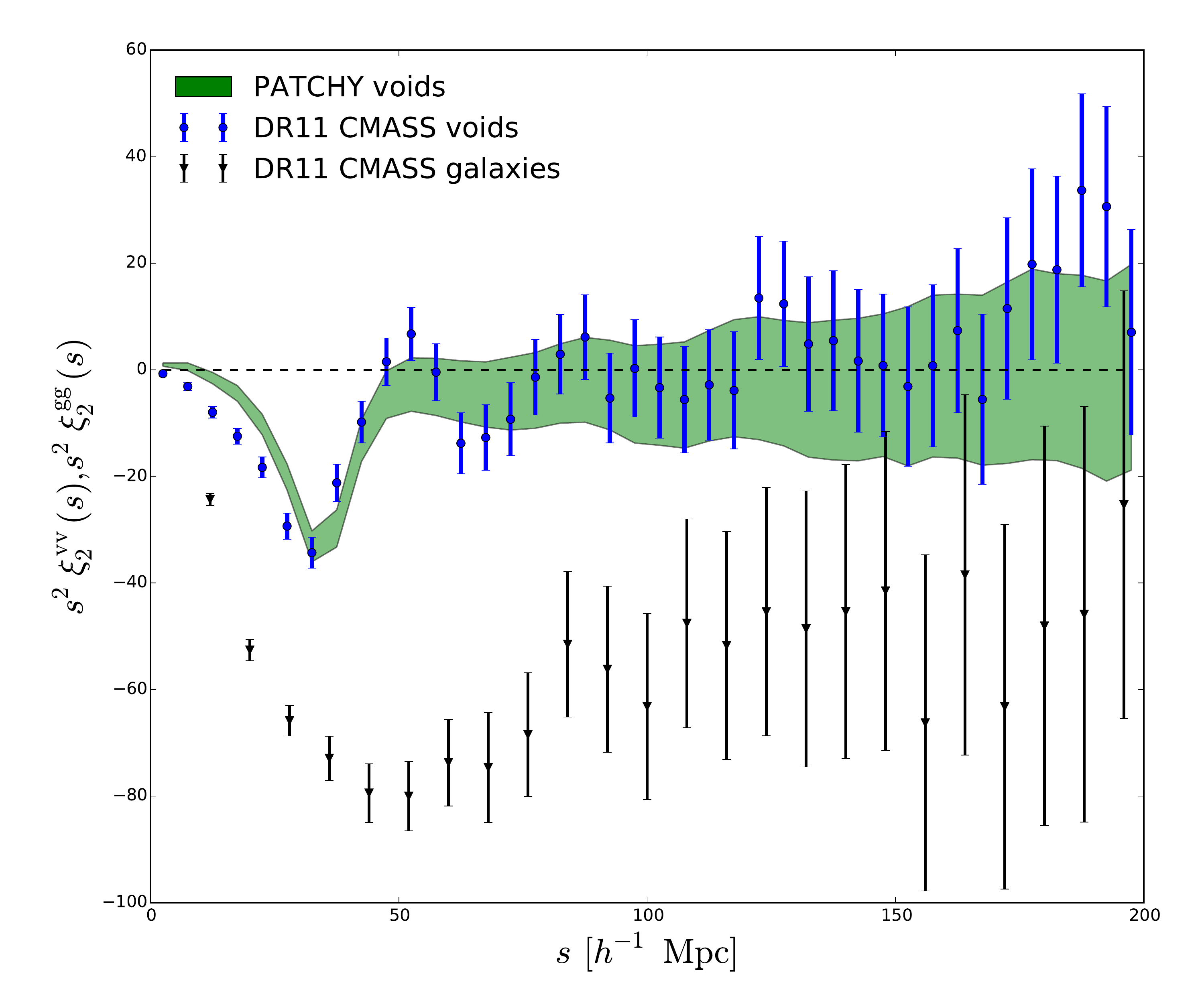}
\caption{Quadrupoles of void correlation functions from CMASS-NGC DR11 voids (blue points) and the averaged correlation function from 1000 \textsc{patchy} mock void catalogs (green area indicates the 1 $\sigma$ region). Black dots and error bars: quadrupole from the CMASS DR11 galaxy clustering. }
\label{fig:quad-cmass}
\end{center}
\efi

%%%%%%%%%%%%%%%%%%%%%%%%%%%%%%%%%%%%%%%%%%%%%%%%%%

% Don't change these lines	% typesetting comment
\label{lastpage}
\end{document}